\pdfoutput=1
\documentclass[preprint,floats,floatfix,aps,12pt]{revtex4}

\usepackage{times}
\usepackage{amssymb}
\usepackage{amsmath}

\ifx\pdfoutput\undefined
\usepackage{graphicx}
\else
\usepackage[pdftex]{graphicx}
\usepackage{epstopdf}
\fi

\usepackage{color}

\newcommand{\mylab}[1]{\label{#1}}
\begin{document}
\title{Front instabilities in evaporatively dewetting nanofluids}
\author{I. Vancea}
\email{vancea@mpipks-dresden.mpg.de}
\author{U. Thiele}
\email{u.thiele@lboro.ac.uk}
\homepage{http://www.uwethiele.de}
\affiliation{Department of Mathematical Sciences, Loughborough University,
Leicestershire LE11 3TU, UK}
\affiliation{Max-Planck-Institut
f\"ur Physik komplexer Systeme, N{\"o}thnitzer Str.\ 38, D-01187 Dresden, Germany}
\author{E. Pauliac-Vaujour}
\author{A. Stannard}
\author{C. P. Martin}
\author{M. O. Blunt}
\author{P. J. Moriarty}
\affiliation{The School of Physics and Astronomy,
        The University of Nottingham,
        Nottingham NG7 2RD, UK}
\begin{abstract}
  Various experimental settings that involve drying solutions or
  suspensions of nanoparticles -- often called nanofluids -- have
  recently been used to produce structured nanoparticle layers. In
  addition to the formation of polygonal networks and spinodal-like
  patterns, the occurrence of branched structures has been
  reported. After reviewing the experimental results we use a modified
  version of the Monte Carlo model first introduced by Rabani et
  al.~[Nature \textbf{426}, 271 (2003)] to study structure formation in
  evaporating films of nanoparticle solutions for the case that all
  structuring is driven by the interplay of evaporating solvent and
  diffusing nanoparticles.  After introducing the model and its
  general behavior we focus on receding dewetting fronts which are
  initially straight but develop a transverse fingering
  instability. We analyze the dependence of the characteristics of the
  resulting branching patterns on the driving chemical potential, the
  mobility and concentration of the nanoparticles, and the interaction
  strength between liquid and nanoparticles. This allows us to
  understand the underlying instability mechanism.
\end{abstract}

\maketitle

%%%%%%%%%%%%%%%%%%%%%%%%%%%%%%%%%%%%%%%%%%%%%%%%%%%%%%%%%%%%%%%%%%%%%%%%%%%%%%%
\section{Introduction}
%%%%%%%%%%%%%%%%%%%%%%%%%%%%%%%%%%%%%%%%%%%%%%%%%%%%%%%%%%%%%%%%%%%%%%%%%%%%%%%
%
Branched structures are ubiquous in nature.  They appear in a wide
variety of phenomena including growing trees, river networks, dendrites
in solidification, and crystallization \cite{Ball99}. In hydrodynamical
systems they occur, for instance, in Saffmann-Taylor fingering
\cite{SaTa58} whose zero-interface tension limit is well described by
the discrete computational model of diffusion limited aggregation
(DLA) \cite{Vics89}.  Branched patterns may also result from contact
line instabilities in spreading drops of surfactant solution
\cite{CaCa99,CACC03,WCM04b}.  This is,
however, rather an exception as in most cases transverse
instabilities of advancing or receding contact lines do not lead to
branched structures. Normally one finds either (i) arrays of straight
parallel or wedge shaped advancing fingers, as for a
gravitationally driven front moving down an incline or vertical wall
\cite{Hupp82,SpHo96,ESR00,Kall00,ThKn03} or driven by thermal
gradients \cite{CHTC90,BBR92,BMFC98}; (ii) fingers which advance radially outwards 
in the case of a spinning drop \cite{MJF89}; or (iii)
fingers \cite{Reit93b,ShRe96,Herm00b} or fields of droplets
\cite{Reit93b,ElLi94,KMHP99,ReSh01} remaining at rest behind a
receding circular or straight dewetting front (cf.~also section~3 of
Ref.~\cite{Thie03}).

For the presently studied system of an evaporating solution of gold
nanoparticles that dewets a silicon substrate
\cite{GSHC99,GeBr00,MTB02,MBM04,Mart07}, fingering was predicted
\cite{YoRa06} using a kinetic Monte Carlo approach
\cite{RRGB03,MBM04,SHR05}. Recent fine-tuned experiments reviewed
below in Section~\ref{exp} have found such instabilities, analyzed
their dependence on the properties of the nanoparticles, and could
furthermore justify the usage of the seemingly simplistic Monte Carlo
model \cite{Paul08}. This will be elucidated in more detail below in
Section~\ref{mod:just}.

In the broader field, dewetting experiments with evaporating
suspensions of particles or macromolecules are performed by various
groups in a wide variety of settings. Examples include drying
macroscopic drops of colloidal solutions of micron- or nanometer-sized
particles \cite{Deeg97,Deeg00,Deeg00b,GRPB04,Huan05}, spin-casted
aqueous collagen solutions \cite{Mert97,Mert98,TMP98}, spin-casted
solutions of polystyrene in benzene \cite{KGMS99}, and spin-casted
solutions of polyacrylic acid (PAA) in toluene \cite{GRDK02}.  The
behavior presented in these studies is generic for a wide class of
solvents and solutes (see, e.g., references in \cite{KGMS99}). More
complex situations have also been studied, such as evaporating thin
films of a binary polymer solution on horizontal \cite{GPBR05} or
inclined \cite{Muel06} substrates, a drying binary suspension of
hard-sphere colloidal particles and a non-adsorbing polymer
\cite{HGP02}, and structuring of polymer solution films caused by
high-temperature evaporation/boiling in a dip-coating setting
\cite{Borm05c,Borm05}.

For the less complex systems, in several regions of the parameter
space, the dewetting behavior of the solutions appears to be very
similar to that described for simple polymeric liquids
\cite{deGe85,Reit92,Thie03,Seem05}. In these regions it is possible to
use the solute to 'image' the dewetting patterns created by the
volatile solvent.  This route was taken, e.g., in
Refs.~\cite{Mert97,TMP98}. This is, however, not the case in other
parameter regions. Most strikingly, the experiments show that dewetting
(evaporating) solutions show much stronger transverse instabilities
of the dewetting front than pure liquids (see, e.g.,
Refs.~\cite{Thie98,GRDK02,Paul08})

The present paper focuses, after a short review of the experiments
(Section~\ref{exp}), on an analysis of the fingering instability
employing the kinetic Monte Carlo model as developed in
Ref.~\cite{RRGB03} and later modified in
\cite{MBM04}. Section~\ref{mod} introduces the model and the numerical
algorithm. Section~\ref{sec:hom} reviews the general behavior of the
model, whereas Section~\ref{sec:finger} entirely focuses on a
parametric analysis of the fingering instability. Finally,
Section~\ref{sec:conc} concludes and gives an outlook of future work.

%%%%%%%%%%%%%%%%%%%%%%%%%%%%%%%%%%%%%%%%%%%%%%%%%%%%%%%%%%%%%%%%%%%%%%%%
\section{Experiments} \mylab{exp}
%%%%%%%%%%%%%%%%%%%%%%%%%%%%%%%%%%%%%%%%%%%%%%%%%%%%%%%%%%%%%%%%%%%%%%%%
%
In the following, we review recent experiments that use monodisperse
colloidal solutions of thiol-passivated gold nanoparticles, sometimes
called a 'nanofluid'.  The 2 or 3\,nm gold core is covered by
alkyl-thiol molecules, whose carbon chain length can be changed from
$C_6$ to $C_{12}$ \cite{Paul08}. Varying the length of the chain
allows control of the interaction parameters, to a certain extent.  In
general, the particles are hydrophobic and are then easily suspended
in toluene.  They can also be made hydrophilic by modifying the
termination of the passivating alkyl-thiol molecules. Then they can be
suspended in water or in another polar solvent. Most experiments are
carried out using hydrophobic particles. Normally, the solution is
deposited (see below) on a silicon substrate that is only covered by
the native silicon oxide layer \cite{MTB02}. However the wetting
behavior of the solvent (and the interaction of the nanoparticles
with the substrate) can be (locally) changed by oxidizing the
substrate further \cite{Mart07}.  The properties of the solvent can be
changed as well, e.g., by adding excess thiol (\emph{which is} thought
to mainly affect the viscosity of the solution \cite{Paul08}).

The actual deposition process is very important to determine
evaporation speed and, in consequence, the results of the ongoing
pattern formation process. Two procedures are followed. On the one
hand, a thin film of solution is deposited by spin-coating a drop of
solution onto the native-oxide terminated silicon substrate.  Then
evaporation competes with dewetting as was described some time ago in
Ref.~\cite{TMP98} for a related system involving a solution of
macromolecules. After all solvent has evaporated the resulting
nanoparticle deposits are imaged using atomic force microscopy
(AFM). Depending on the concentration of the solution one may find
mono-modal or bi-modal cellular network structures as in
Ref.~\cite{MTB02}, ribbon-like or labyrinthine structures, or branched
structures (see Fig.~\ref{fig:nets}). Flower-like branched structures
are also observed (Fig.~\ref{fig:flowers}).  Note that for spin-coated
films the evaporation is fast and normally the structuring is complete
even before the spin-coater is stopped. At present, structuring of the
film under spin-coating conditions has not been observed \emph{in
  situ}.

\begin{figure*}[htbp]
\includegraphics[width=0.95\hsize]{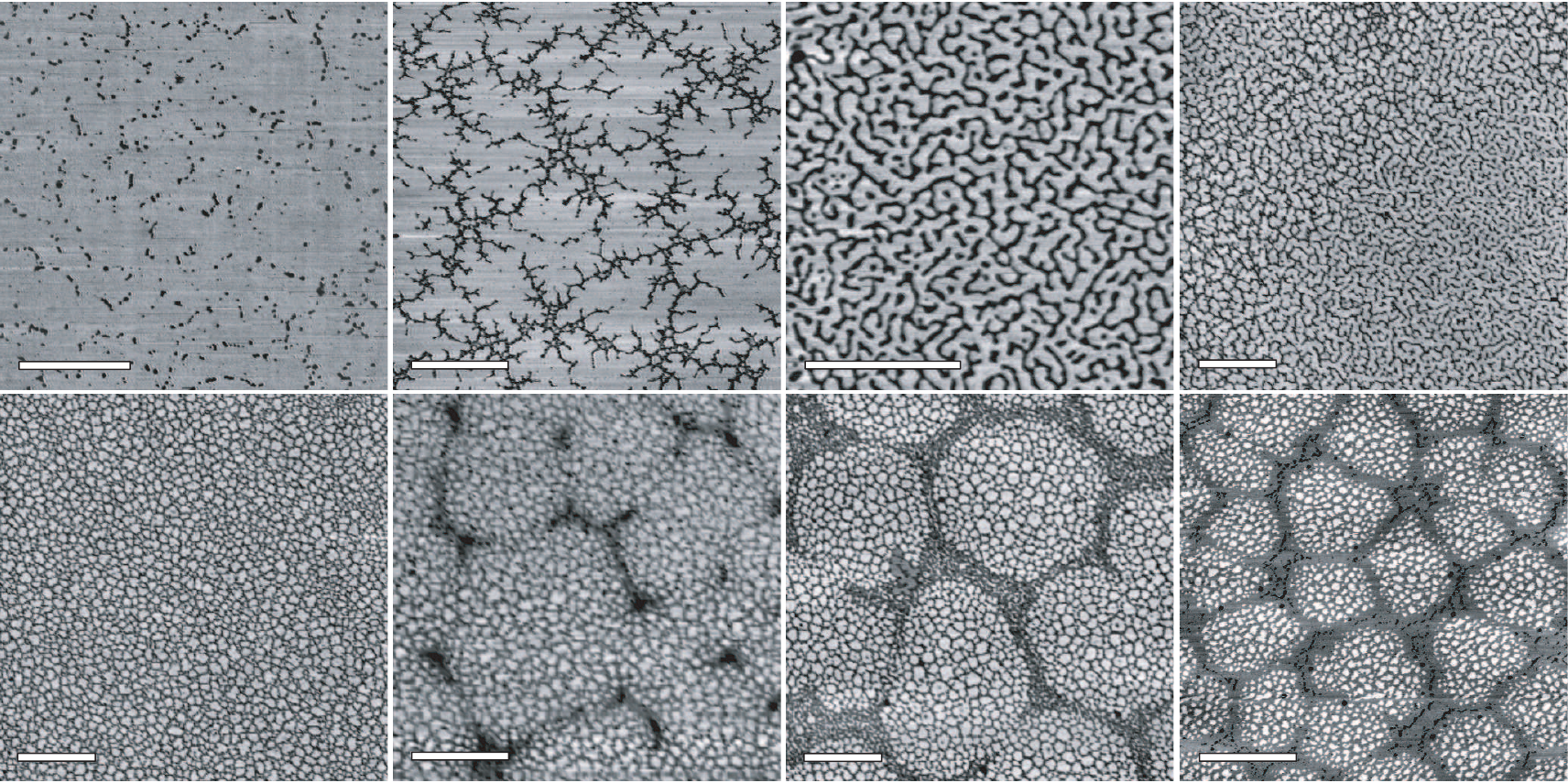}
\caption{AFM images of dodecanthiol-passivated 2nm gold nanocrystals,
  spin-coated from toluene onto native oxide
  terminated silicon substrates.  From top left to bottom right,
  concentrations increase as follows: 0.10mg/ml, 0.25mg/ml, 0.50mg/ml,
  0.75mg/ml, 1.00mg/ml, 1.25mg/ml, 1.50mg/ml and 1.75mg/ml. Scans are
  $2\times2$ to $5\times5\mu$m$^2$ in size. All scale bars correspond
  to $1\mu$m. The empty substrate is white, the deposited
  nanoparticles are black.}
\mylab{fig:nets}
\end{figure*}

\begin{figure}[htbp]
(a)\includegraphics[width=0.9\hsize]{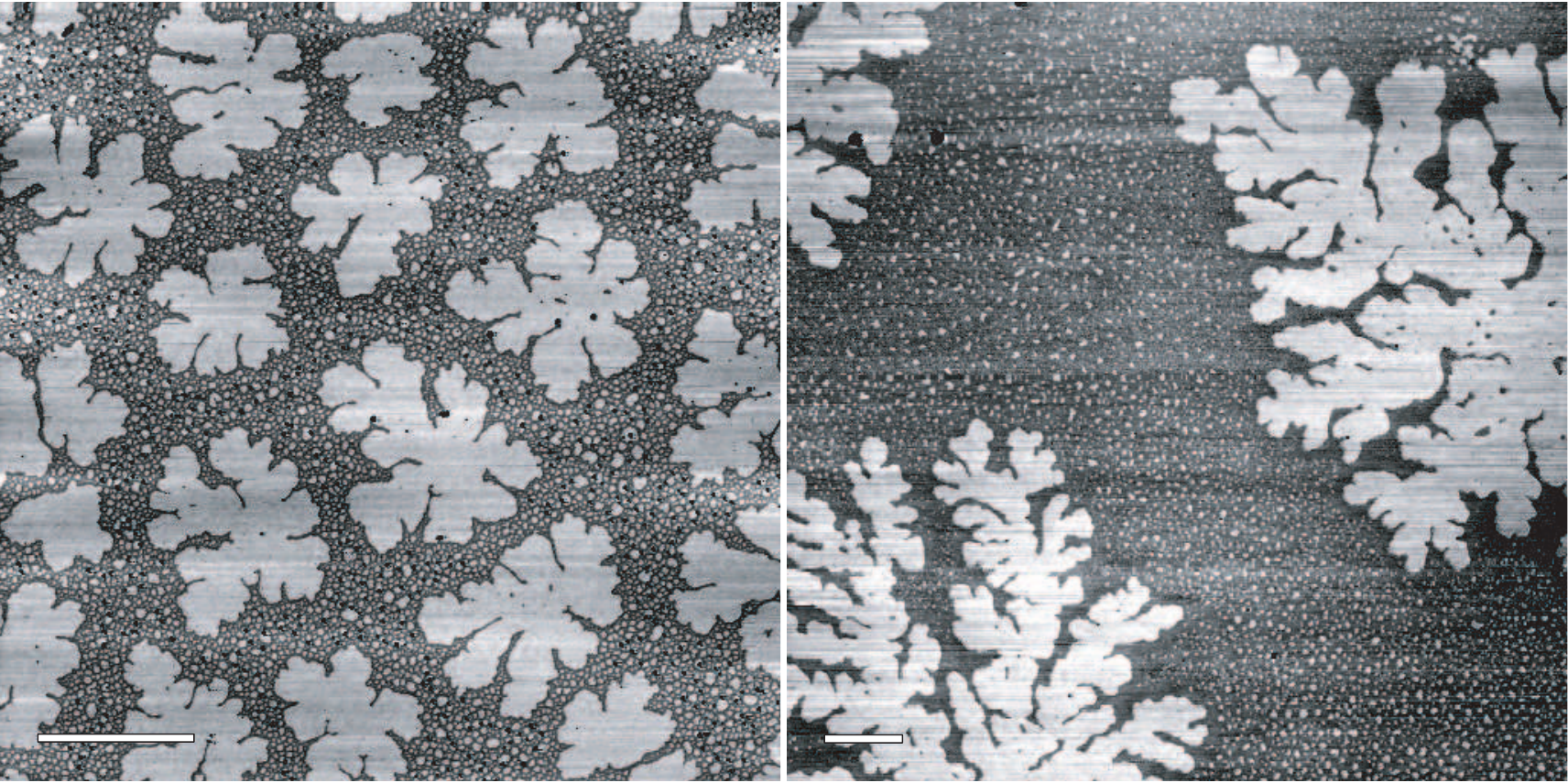}(b)
\caption{AFM pictures of flower-like branched structures in the case
  of (a) a toluene solution of gold nanoparticles spin-coated onto
  native oxide terminated silicon substrates, and (b) a toluene
  solution (with an excess of thiol) deposited using the meniscus
  technique. The empty substrate is white, the deposited nanoparticles
  are black. Scale bars correspond to $2\mu$m.}
\mylab{fig:flowers}
\end{figure}

A meniscus technique using a teflon ring was recently
introduced \cite{PaMo07} to decrease the solvent evaporation rate
 during nanoparticle self-organization. Precursors of
this technique have involved the use of latex beads (see, e.g., Ref.~\cite{GDD01}
and references in Ref.~\cite{PaMo07}).  The slowing-down of the
evaporation process improves control and allows the use of contrast-enhanced
microscopy \cite{nanolane} to capture the dewetting process on video
\cite{Paul08}. In the meniscus technique a drop of solution is
deposited onto a teflon ring that sits on the silicon substrate. As
the toluene wets teflon, it evaporates quickly (within seconds) from
the center of the ring only. The meniscus thereby formed between the
silicon substrate and the teflon ring slowly shrinks due to
evaporation on the time scale of one hour. The pattern formation
process is confined to the region of the receding toluene/silicon/air
contact line.

\begin{figure}[htbp]
\includegraphics[width=0.7\hsize]{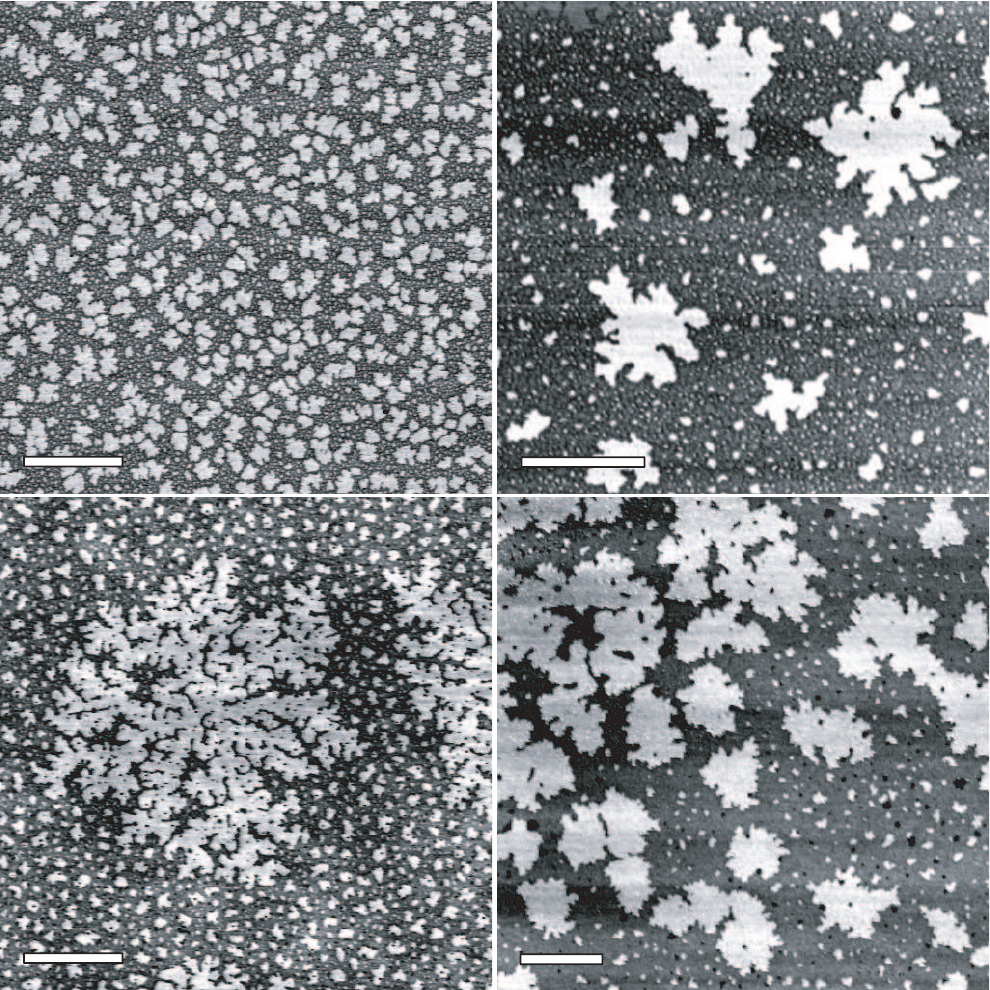}
\caption{AFM pictures of fingering patterns obtained by evaporative
  dewetting using the meniscus technique. The employed
  thiol-passivated gold nanoparticles are coated with thiol molecules
  of different lengths. Shown are images representative of chain
  length (a) $C_8$, (b) $C_{10}$, (c) $C_{12}$, and (d) $C_{14}$. No excess
  thiol is present. The empty substrate is white, the deposited
  nanoparticles are black. Scale bars correspond to $1\mu$m.}
\mylab{fig:finger}
\end{figure}

Using the meniscus technique one can observe (as for the spin-coating
procedure) labyrinthine spinodal structures and mono-modal or bi-modal
network patterns depending on nanoparticle concentration and position
inside the teflon ring. (The position within the ring is closely
linked to the solvent evaporation rate). It therefore becomes possible
to study branched structures in a more controlled manner.  Selected
examples of such structures are given in Fig.~\ref{fig:finger}. The
systematic study performed in Ref.~\cite{Paul08} indicates two
important effects: Fingering strongly depends (i) on the chain length
of the thiol molecules passivating the gold core of the particles and
it also depends (ii) on the amount of excess thiol in the solution.
The chain length is varied from pentane (C$_5$) to tetradecane
(C$_{14}$) at fixed nano\-particle concentration and deposited volume
[see Figs.~2(a)-(e) of Ref.~\cite{Paul08}]. For C$_5$ and C$_8$
ligands, no formation of branched structures is observed irrespective
of evaporation rate, i.e., position within the teflon ring. For longer
chains (C$_{10}$ and C$_{12}$) well-developed branched structures are
formed.  Increasing the chain length even further (C$_{14}$), however,
produces less developed branching.  An addition of a small amount of
excess thiol to the solution (0.1\% by volume) enhances the
development of branched structures strongly [see Ref.~\cite{Paul08}].
For small chain length' (C$_5$- and C$_8$) branching (albeit sometimes
weak) is now observed.  The branching is considerably stronger for
C$_{10}$, C$_{12}$ and C$_{14}$.

The main aim of the present work is to shed more light on the
mechanisms underlying the branching process using a simple
computational model \cite{RRGB03,MBM04}. To separate the process from
other effects we Analise the dependence of the branching of a straight
evaporative dewetting front on the control parameters. The results are
used to explain the experimental findings and indicate avenues for
future experiments. We will come back to this below in the Conclusion.

%%%%%%%%%%%%%%%%%%%%%%%%%%%%%%%%%%%%%%%%%%%%%%%%%%%%%%%%%%%%%%%%%%%%%%%%
\section{Kinetic Monte Carlo model}
\mylab{mod}
%%%%%%%%%%%%%%%%%%%%%%%%%%%%%%%%%%%%%%%%%%%%%%%%%%%%%%%%%%%%%%%%%%%%%%%%
\subsection{Justification of usage}
\mylab{mod:just}
%%%%%%%%%%%%%%%%%%%%%%%%%%%%%%%%%%%%%%%%%%%%%%%%%%%%%%%%%%%%%%%%%%%%%%%%

A simple two-dimensional kinetic Monte Carlo model for the present
process was first proposed in Ref.~\cite{RRGB03} to model experiments
such as those presented in Ref.~\cite{GeBr00}.  Before we introduce
the model in detail, we first discus its applicability and justify its
usage. The model is based on two key assumptions: (1) that the relevant
processes can be captured by a two-dimensional setting neglecting
changes in the film thickness of the evaporating film; and (2) that
all relevant dynamics are a result of the diffusion of nanoparticles and
of the evaporation of the solvent. Convective motion of the solution
is entirely neglected. A refined model increases the spatial range of
considered energetic interactions including next-nearest neighbors
\cite{MBM04} but otherwise uses the same basic assumptions.
Considering the strong assumptions, the agreement with experiments is
amazingly good \cite{RRGB03,MBM04}. As we shall now discuss, recent
experiments using the meniscus technique (described above) may explain
why.

In Ref.~\cite{Paul08} the evolution of the branched patterns is
followed in real-time using contrast-enhanced video microscopy. The
video (complementary material of Ref.~\cite{Paul08}) clearly shows
that different processes occur on different scales. First, a macroscopic
dewetting front recedes, leaving behind a seemingly dry substrate. The
macroscopic front can be transversally unstable resulting in
large-scale ($>100\mu$m) strongly anisotropic finger structures. For
fronts that move relatively quickly these macroscopic structures cover
all available substrate. However, when at a later stage the
macroscopic front becomes slower, those fingers become scarce and
'macroscopic fingering' finally ceases. At this stage it is possible
to appreciate that the seemingly dry region left behind by the front
is not at all dry, but covered by an ultrathin 'postcursor' film that
is itself not stable.  At a certain distance from the macroscopic
front the ultrathin film starts to evolve a locally isotropic pattern
of holes. The holes themselves grow isotropically in an unstable
manner resulting in an array of isotropically branched structures as
shown, e.g., above in Figs.~\ref{fig:flowers} and
\ref{fig:finger}. This indicates that {\it nearly all} of the patterns
described in various publications result from processes in the
ultrathin film whose thickness is of the order of the size of the
nanoparticles.

Focusing on the structuring of the ultrathin film, we next identify
the important dynamical processes.  On the mesoscopic scale the
dynamics of a thin film of pure liquid can be described by a thin film
equation derived using a long wave approximation
\cite{ODB97,KaTh07}. In such a model, the temporal change in film
thickness $h$ results from the gradient of the convective flow $h^3
(\nabla p) / 3 \eta$ and from an evaporative loss $\beta(\rho\mu_0-
p)/\rho$ \cite{LGP02,Pism04}. Here, $\eta$ and $\rho$ are the dynamic
viscosity and density of the solvent, respectively, $\mu_0$ is the
chemical potential (related to the ambient vapor pressure of the
solvent), and $\beta$ is a rate constant that can be obtained from gas
kinetic theory or from experiment \cite{LGP02}.  The pressure $p$
contains curvature and disjoining pressures and drives {\it both}
processes -- convection and evaporation.  The disjoining pressure
describes the wettability of the substrate
\cite{deGe85,Isra92,Mitl93,Thie03}. Due to the thickness dependence
$h^3$, the mobility related to convective motion is large for thick
films but decreases strongly with decreasing film thickness whereas
the mobility related to evaporation (the rate constant) is a constant.
In consequence, it is expected that for a nanometric film the
evaporation term strongly dominates. Using the parameters given in
Ref.~\cite{LGP02} and assuming a small contact angle of about 0.01,
one obtains a cross-over thickness in the lower single digit nanometer
range. Below this thickness the solvent dynamics is dominated by
evaporation.

This consideration, together with the experimental observation
discussed above, justifies the neglect of convective motion in the
kinetic Monte Carlo model. Without convection a two-dimensional model
is sufficient to cover the essential processes.  Note that a
refinement proposed in \cite{Mart07} introduces a three-dimensional
aspect into the two-dimensional model by making the chemical potential
dependent on mean liquid coverage (i.e., on a parameter related to
mean thickness). This amounts to a consideration of the
thickness-dependent disjoining pressure in the evaporation term {\it
  without} the explicit incorporation of a film thickness.  The
resulting pseudo three-dimensional model successfully reproduces
bi-modal structures\cite{Mart07}. These shall, however, be of no
concern here.

\subsection{The model}

Kinetic Monte Carlo models are a common tool to investigate a
wide variety of dynamical processes \cite{ChVl07,FW91}. In particular,
they have proven to be attractive tools to model adsorption,
diffusion, and aggregation in surface science resulting in interfacial
growth and structuring like, e.g., the growth of metal layers as atoms
arrange after being deposited on surfaces \cite{KW89,L01,RB90}.

\begin{figure}[htbp]
\includegraphics[width=8cm]{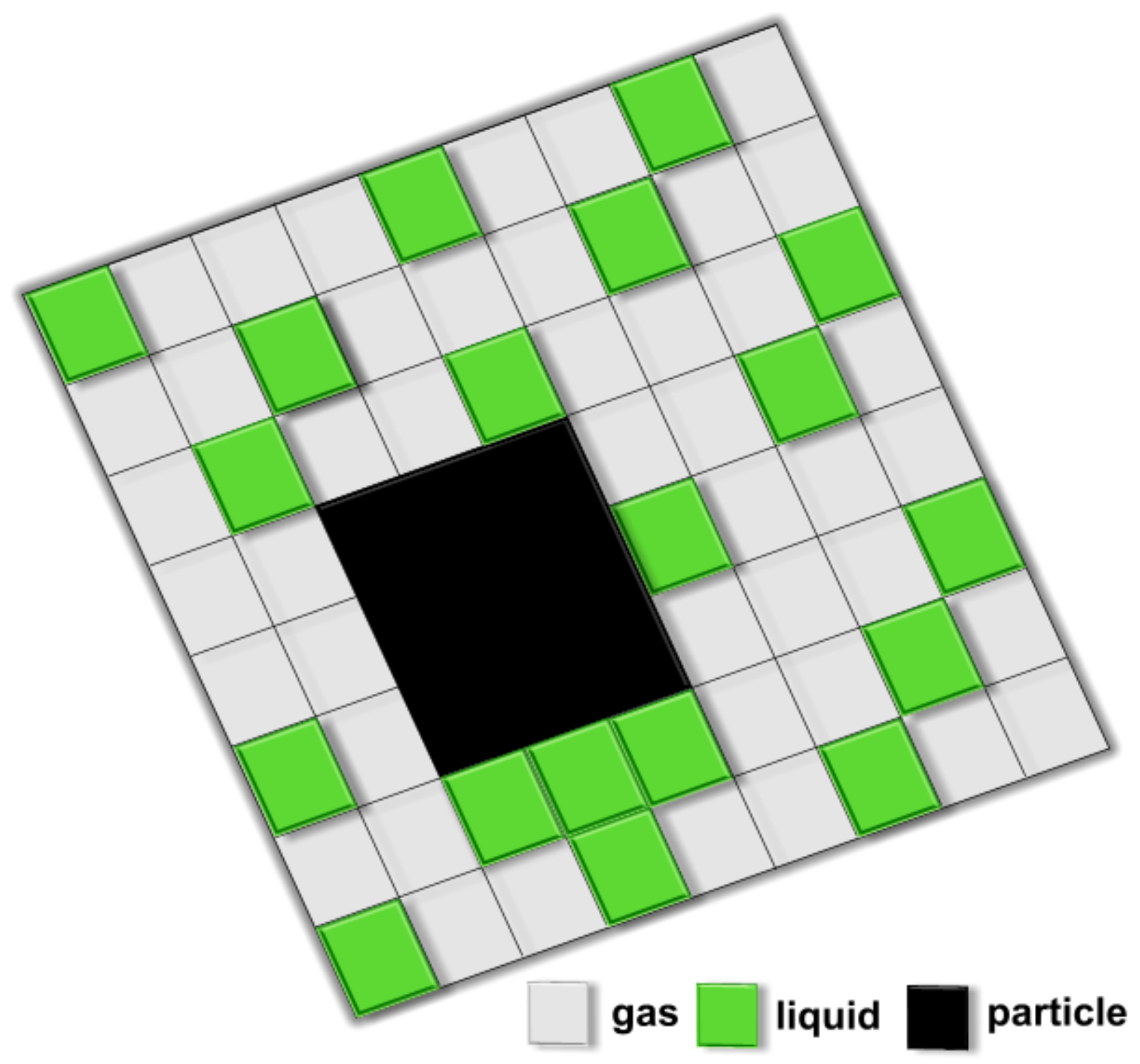}
\caption{Schematic picture of the two-dimensional lattice gas model
used: the size of nanoparticle is in our case $3\times3$, whereas a 
liquid cell is of size $1\times1$. The side length of such a cell
represents the correlation length of the solvent approximated to be
1\,nm \cite{RRGB03}.}
\mylab{fig:sketch}
\end{figure}

The approach followed in Ref.~\cite{RRGB03} to model an evaporating
dewetting nanoparticle solution is based on an Ising-type model for
the liquid-gas phase transition.  To facilitate comparison with
previous work we here choose the model to be exactly the same as in
Ref.~\cite{MBM04} (building on Ref.~\cite{RRGB03}): The system is
described using a two-dimensional lattice gas of two fields $n$ and
$l$.  We allow for three types of possible states of a cell -- liquid
($l=1, n=0$), nanoparticle ($l=0, n=1$), and vapor ($l=0, n=0$, i.e.,
cell empty). Thereby liquid/vapor is assigned per $1\times1$ cell
whereas it is assumed that the nanoparticles fill $3\times3$ cells
(see sketch Fig.~\ref{fig:sketch}) to indicate the larger size of the
nanoparticles used in the experiment as compared to the correlation
length of the liquid.  This detail, however, turns out to be not very
relevant as further discussed in the Conclusion.

The energy of a configuration of the two-dimensional assembly of
vapor, liquid and nanoparticles is determined by the Hamiltonian
\begin{equation}
E\,=\,-\frac{\varepsilon_{nn}}{2}\sum_{<ij>} n_i n_j
\,-\,\frac{\varepsilon_{nl}}{2}\sum_{<ij>} n_i l_j
\,-\,\frac{\varepsilon_{ll}}{2}\sum_{<ij>} l_i l_j
\,-\,\mu\sum_{i} l_i
\mylab{eq:ham}
\end{equation}
where $\varepsilon_{ll}$, $\varepsilon_{nn}$ and $\varepsilon_{nl}$
are the interaction energies for adjacent sites $(i,j)$ filled by
(liquid,liquid), (nanoparticle,nanoparticle) and
(liquid,nanoparticle), respectively.  For fixed interaction
strengths, the equilibrium state is determined by the chemical
potential $\mu$ that controls the evaporation/condensation of the
liquid. We will call $|\mu|$ the driving force as it is primarily
responsible for the motion of a dewetting or wetting front.  The sums
$\sum_{<ij>}$ are taken over all pairs of nearest and next-nearest
neighbors.  Thereby, the interaction strength of the next nearest
neighbors is corrected by a factor $1/\sqrt{2}$ due to their larger
distance \cite{MBM04}. In the following we fix $\varepsilon_{ll}=1$,
i.e., we express all energies and the chemical potential in the scale
of the liquid-liquid interaction energy.

The energy functional determines the equilibrium state and the energy
landscape of the system. The dynamics is determined by the allowed
Monte Carlo moves, their relative frequency, and the rules for their
acceptance.  Two types of moves are allowed: (i)
evaporation/condensation of liquid and (ii) diffusion of
nanoparticles within the liquid. To simulate the dynamics, the energy loss
or gain $\Delta E$ related to a potential move is calculated.
The move is then accepted with the probability $p_{\mathrm{acc}} = \min[1,
\exp(-\Delta E/kT)]$ where $k$ is the Boltzmann constant and $T$ the
temperature. The rule implies that any move that decreases the energy
of the system is accepted when tried.  Any move that increases the
energy has a probability corresponding to the related Boltzmann
factor. Consistent with the above scales, temperature is expressed in
units of $\varepsilon_{ll}=1/k$. It determines the importance of
fluctuations in the system. For $T=0$ the system is fluctuation-free,
i.e.\ the evolution follows a deterministic gradient dynamics.

Practically, we use a checkerboard Metropolis algorithm to advance the
liquid/vapor subsystem; each solvent or vapor cell is examined in
turn, and is converted from liquid to vapor or from vapor to liquid
with the acceptance probability $p_{\mathrm{acc}}$. After one solvent
cycle all particles are considered and a diffusive move is tried.  The
nanoparticles are free to perform a restricted random walk on the
lattice. It is restricted because particles are only allowed to move into
wet areas of the substrate, i.e., onto cells with $l=1$. This is a
very important rule and models zero diffusivity of the
particles on a dry substrate. If such a 'forbidden' move is tried,
the particle is left at rest and the next particle is considered.  The
diffusivity or mobility of the nanoparticles inside the liquid is
controlled by the number of times each particle is examined after one
solvent cycle.  This computational ratio $M$ of particle and
solvent cycles reflects the physical ratio of time scales for
evaporation and diffusion. Large $M$, for instance, indicates that
diffusion is fast as compared to evaporation, i.e., it stands for a
large diffusion constant of the nanoparticles or/and a low
evaporation rate for the liquid. It therefore represents a low
effective viscosity of the solution.

The model is used in the following to first review some general
results for a homogeneous system without and with nanoparticles
(Section~\ref{sec:hom}). Then we focus on a parametric analysis of
the fingering instability (Section~\ref{sec:finger}).

%%%%%%%%%%%%%%%%%%%%%%%%%%%%%%%%%%%%%%%%%%%%%%%%%%%%%%%%%%%%%%%%%%%%%%%%
\section{Behavior of a homogeneous system} \mylab{sec:hom}
%%%%%%%%%%%%%%%%%%%%%%%%%%%%%%%%%%%%%%%%%%%%%%%%%%%%%%%%%%%%%%%%%%%%%%%%
%
\subsection{Without nanoparticles}
\mylab{sec:won}
Without nanoparticles the behavior of the model introduced in
Section~\ref{mod} is well known, it reduces to the classical Ising
model in two dimensions describing the order-disorder transition in a
ferromagnetic system.  Using the mapping $\mu \rightarrow \mu_0 H - 2$
and $l \rightarrow (s+1)/2$ where $s=\pm1/2$ corresponds to the spin
of a cell, $\mu_0$ to the magnetic moment and $H$ to the external
magnetic field, all results since Onsager \cite{Onsa1944} can be
used. In particular, an infinitely extended system has a critical
point at $\mu_c=-2$ and $kT_c^\infty=1/[2
\ln(1+\sqrt{2})]\approx0.567$. For temperatures below $T_c^\infty$ at
$\mu_{\rm ph}=-2$ the liquid state and the vapor state may coexist,
whereas for $\mu<-2$ [$\mu>-2$] eventually the vapor [liquid]
state dominates. A chemical potential not equal to -2 corresponds to a
non-zero external magnetic field in the ferromagnetic system.

For $T<T_c^\infty$ there exists a first order phase transition at
$\mu_{\rm ph}$; systems in the liquid [gas] state are metastable for a
certain range below [above] $\mu_{\rm ph}$. Mean field theory as
presented, e.g., in Ref.~\cite{Lang92} provides the line that is the
lower [upper] limit for the existence of metastable liquid [gas]
states in the $(\mu, kT)$ plane. The two curves fulfill
\begin{equation}
  \mu_{\rm ms}^\pm \,=\,\pm\,\frac{2}{3}\,kT\left(\frac{T_c^\infty}{T}-1\right)^{3/2} \,-\,2,
  \mylab{eq:spin}
\end{equation}
where we set the grid spacing used in Ref.~\cite{Lang92} to one.

\begin{figure}[htbp]
\includegraphics[width=0.9\hsize]{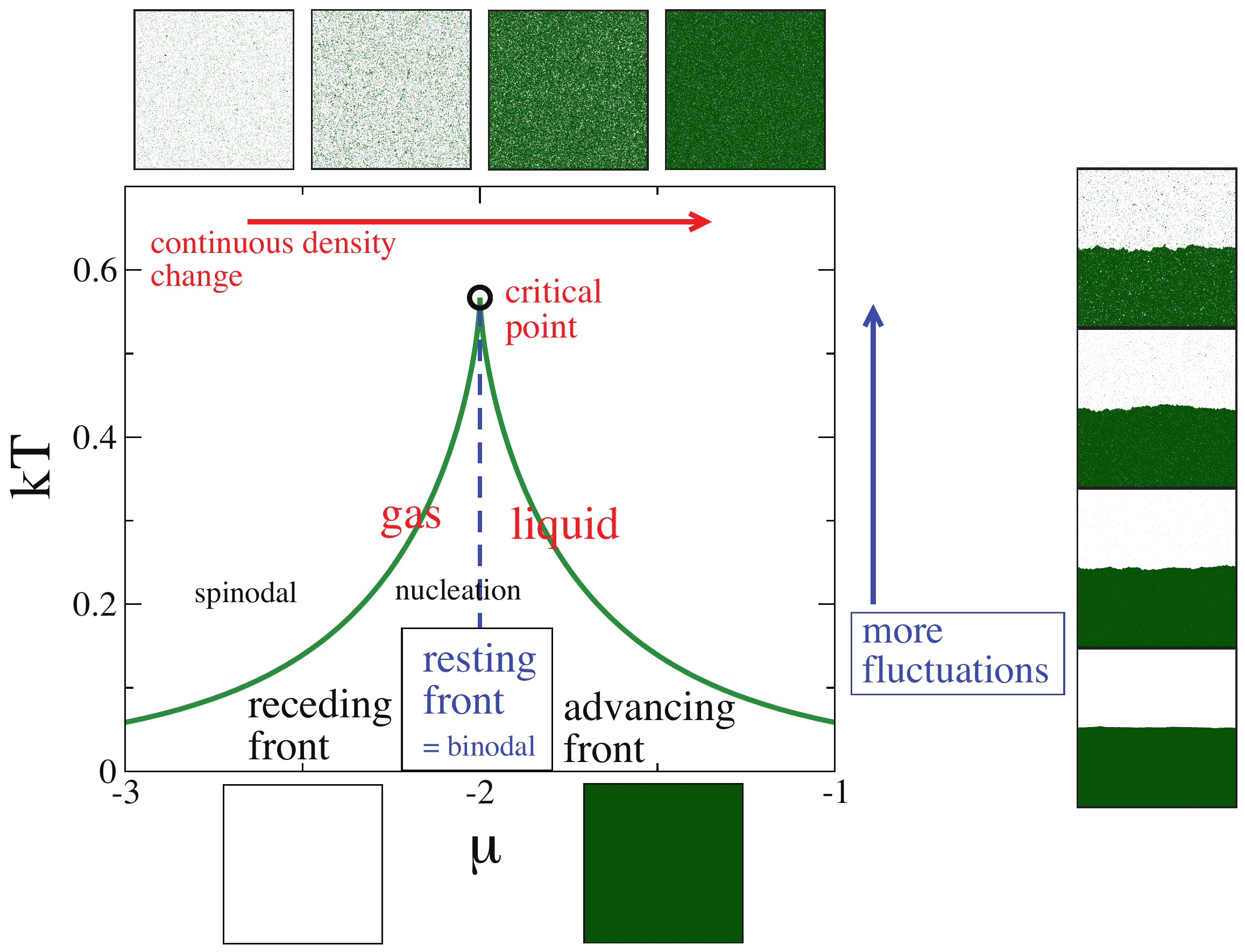}
\caption{Phase diagram without nanoparticles (central panel). A
  schematic overview of system behavior is given in the plane spanned
  by the dimensionless chemical potential $\mu$ and the dimensionless
  measure of temperature $kT$.  The liquid and gas phase are
  indistinguishable above the critical temperature $kT_c=1/[2
  \ln(1+\sqrt{2})]\approx0.57$. The corresponding gradual change of
  mean density with $\mu$ is indicated by the four snapshots of
  equilibrium states shown in the upper line (from left to right: $\mu
  = -2.75, -2.25, -1.75$ and $-1.25$; $kT=0.8$).  Below $T_c$ the
  system can show a rich dynamics when evolving towards equilibrium.
  For $\mu>-2$ [$\mu<-2$] the equilibrium corresponds to liquid [gas]
  as illustrated by the snapshots in the bottom row ($\mu=-1.5$
  [$\mu=-2.5$] and $kT=0.05$), whereas at $\mu_{\rm ph}=-2$ gas and
  liquid may coexist. At $\mu_{\rm ph}$ a straight front separating
  gas and liquid does not move on average. However, depending on
  temperature it fluctuates, as illustrated by the snapshots in the
  right column (from bottom to top $kT = 0.1, 0.3, 0.45$ and $0.55$;
  $\mu=\mu_{\rm ph}=-2$). A straight liquid front will recede
  [advance] for $\mu<-2$ [$\mu>-2$], i.e.~ one is able to study
  evaporative dewetting [wetting] fronts.  Starting, however, with an
  homogeneous liquid-covered [gas-covered] substrate for $\mu<-2$
  [$\mu>-2$] the substrate will empty [fill] via a nucleation or
  spinodal-like process. The borders between the different processes
  are indicated as heavy lines and are discussed in the main text.
  All snapshots are obtained from simulations for a small domain size of
  $300\times300$ after 1000 kinetic Monte Carlo steps.  }
  \mylab{fig:phase}
\end{figure}

An overview of the equilibrium phases and the resulting behavior of a
straight wetting/dewetting front (as described next) is given in
Fig.~\ref{fig:phase}.  Starting with a liquid covered substrate in a
region ($\mu<-2$) where the global energy minimum corresponds to the
vapor phase leads to an evaporative dewetting process that either
follows a nucleation and growth pathway or results from a
spinodal-like process. Typical snapshots illustrating the two
processes are shown in Fig.~\ref{fig:withoutnanopart}.  Nucleation and
growth of holes occurs in the parameter region close to $\mu_{\rm
  ph}=-2$ where the homogeneous liquid state is still metastable. At
smaller $\mu<\mu_{\rm ms}^-$, i.e., at larger driving forces $|\mu|$ a
spinodal-like process occurs when starting with a liquid-filled
plane. Many very small holes appear at once, and all liquid evaporates
very quickly, leaving the substrate empty. When
starting with an homogeneous gas state at $\mu>-2$, similar mechanisms
result in a liquid-filled plane. The relevant border there is
$\mu=\mu_{\rm ms}^+$.

\begin{figure}[htbp]
\centering
(a)\includegraphics[width=0.88\hsize]{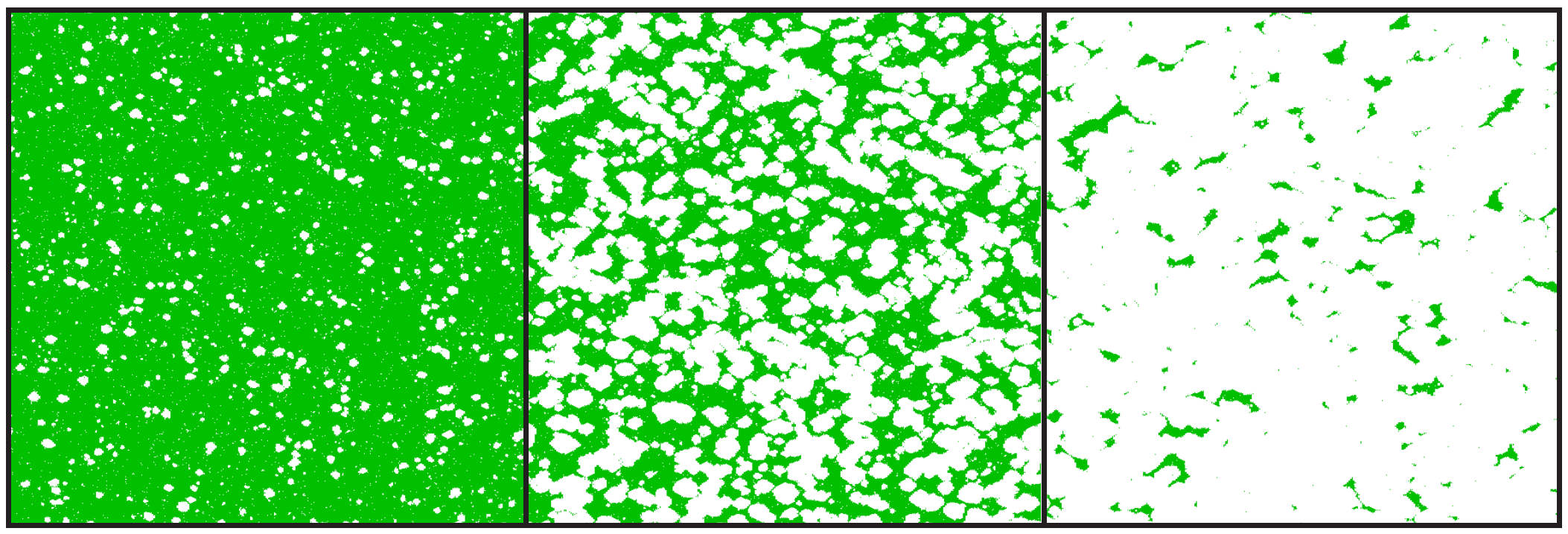}\\
(b)\includegraphics[width=0.88\hsize]{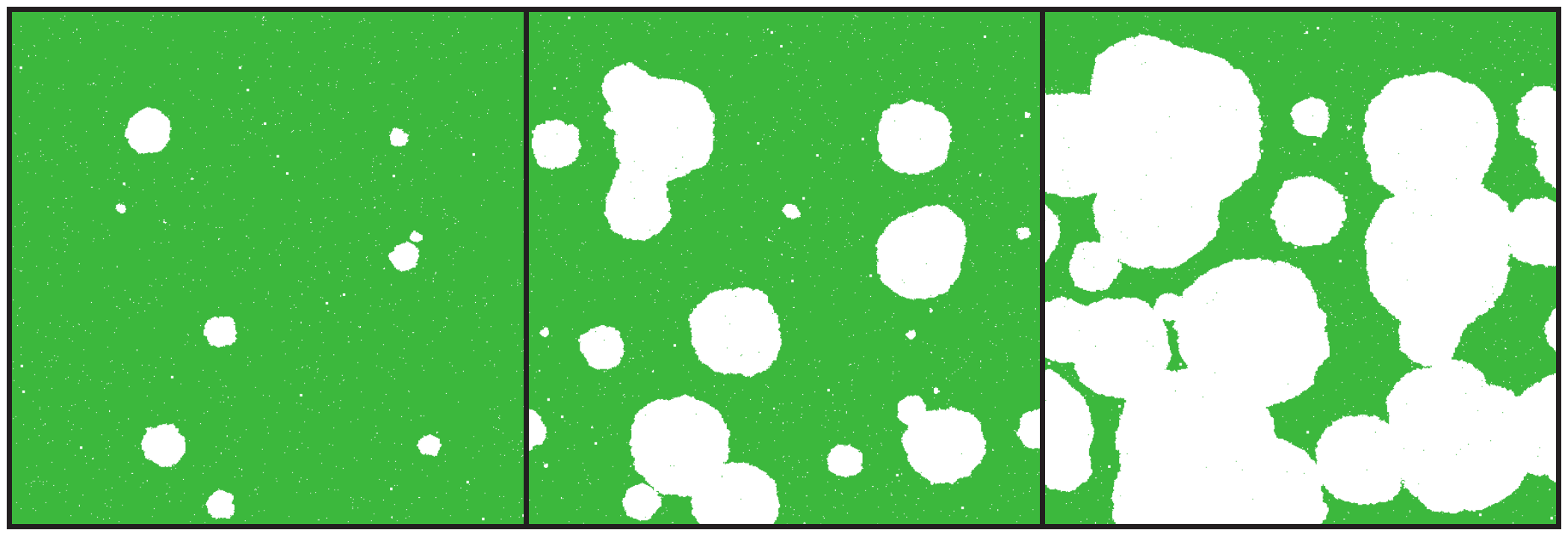}
\caption{Snapshots from typical evaporative dewetting processes
  without nanoparticles in the case of (a) a spinodal-like process at
  $\mu=-2.55$ and (b) of nucleation and growth of holes at $\mu=-2.3$.
  The starting condition in each case is an homogeneous liquid
  film. The number of Monte Carlo steps is from left to right (a): 5,
  10, and 15; (b): 40, 80, and 120; $kT=0.3$ and the lattice size is
  $600\times600$. Liquid is gray (green online) and the empty
  substrate is white.}
\mylab{fig:withoutnanopart}
\end{figure}

A straight front separating vapor- and liquid-covered substrate areas
remains on average at rest for $\mu=-2$ although it might fluctuate
quite strongly. The liquid state will advance and recede for $\mu>-2$
and $\mu<-2$, respectively.  For a resting or moving front the
temperature ($T<T_c$) determines the strength of the fluctuations that
modulate the straight front (see snapshots at the right of
Fig.~\ref{fig:phase}).  The overall picture is similar for circular
fronts, however, the exact value of $\mu$ that results in a resting
front depends on the average curvature of the front (i.e.~the size of
the hole).  A two-dimensional 'curvature pressure' enters the
balance. Fixing some $\mu_{\rm ms}^-<\mu<-2$, a hole in a liquid layer
will grow [shrink] above [below] a certain radius $r_c(\mu)$. The
growing hole will remain almost circular. Practically, on a square
grid in late stages it will become square-like (original model in
Ref.~\cite{RRGB03} where only nearest neighbors are
considered). Including next nearest neighbors leaves the holes
circular up to a later stage, when they become octagonal.

%%%%%%%%%%%%%%%%%%%%%%%%%%%%%%%%%%%%%%%%%%%%%%%%%%%%%%%%%%%%%%%%%%%%%%%%
\subsection{With nanoparticles}
\mylab{sec:wn}
%%%%%%%%%%%%%%%%%%%%%%%%%%%%%%%%%%%%%%%%%%%%%%%%%%%%%%%%%%%%%%%%%%%%%%%%%
%
%
Having discussed the behavior of the model without particles we now
turn our attention towards situations where nanoparticles are
present. First, we discuss the equilibrium behavior, followed by an
analysis of the evolution of an initially liquid-filled well-mixed
homogeneous system.

We expect the particles to influence the location of the liquid-gas
phase transition. Inspecting the Hamiltonian (\ref{eq:ham}) one can
estimate the influence using a mean field argument.  Let us consider a
liquid-filled cell at a straight liquid front.  We consider the energy
needed to move the front, i.e., to empty the cell.  On average a cell
at a straight front has two direct neighbors occupied either by liquid
or nanoparticles. We replace the nanoparticle occupation number in the
liquid-particle interaction term by its mean value, i.e., the coverage
$\phi$, neglect the particle-particle interaction term as we consider
a cell either filled by liquid or gas, and replace one of the liquid
occupation numbers in the liquid-liquid interaction term by its mean
value for a filled cell $1-\phi$.
In this way we can map the model onto the pure liquid-gas case
by replacing $\mu$ by $\tilde{\mu}=\mu+2(\varepsilon_{nl}-1)\phi$.
Following this argument
we expect the phase transition to occur at $\tilde{\mu}=-2$, i.e., an
attractive liquid-gas interaction ($\varepsilon_{nl}>0$) stronger than
the liquid-liquid interaction ($\varepsilon_{ll}=1$) implies that the
phase transition occurs at $\tilde{\mu}_{\rm
  ph}=-2-(\varepsilon_{nl}-1)\phi< \mu_{\rm ph}$. Note, however, that for low
particle concentrations (e.g., $\phi=0.1$) the change is rather small for
$\varepsilon_{nl}=1.5$, i.e. the value used in most of the present work.
This agrees with simulations where we have not spotted a significant shift for the used
concentrations $\le0.2$.

\begin{figure}[htbp]
\centering
(a)\includegraphics[width=0.88\hsize]{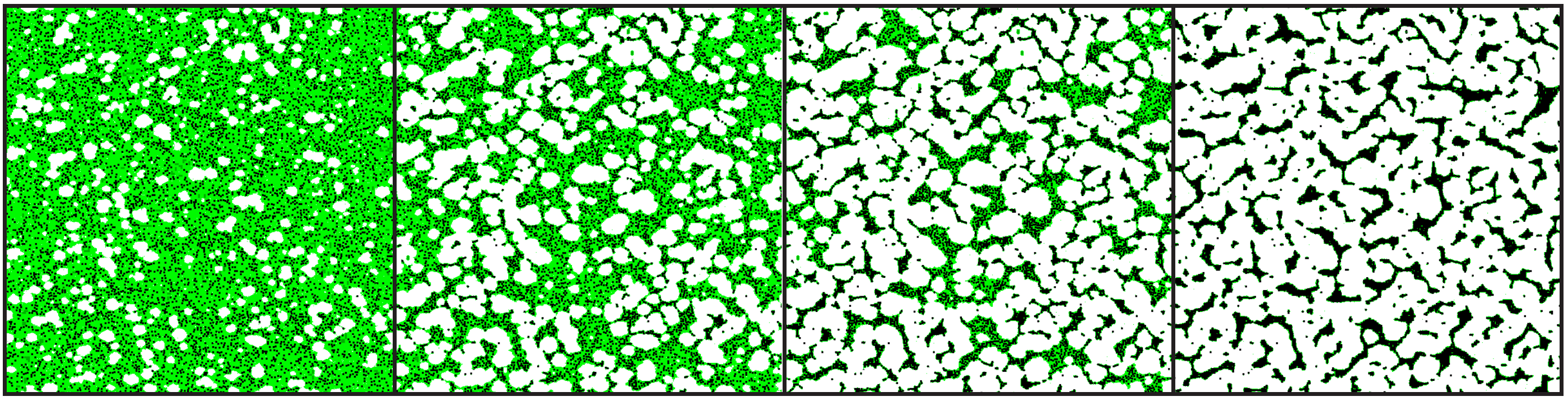}\\
(b)\includegraphics[width=0.88\hsize]{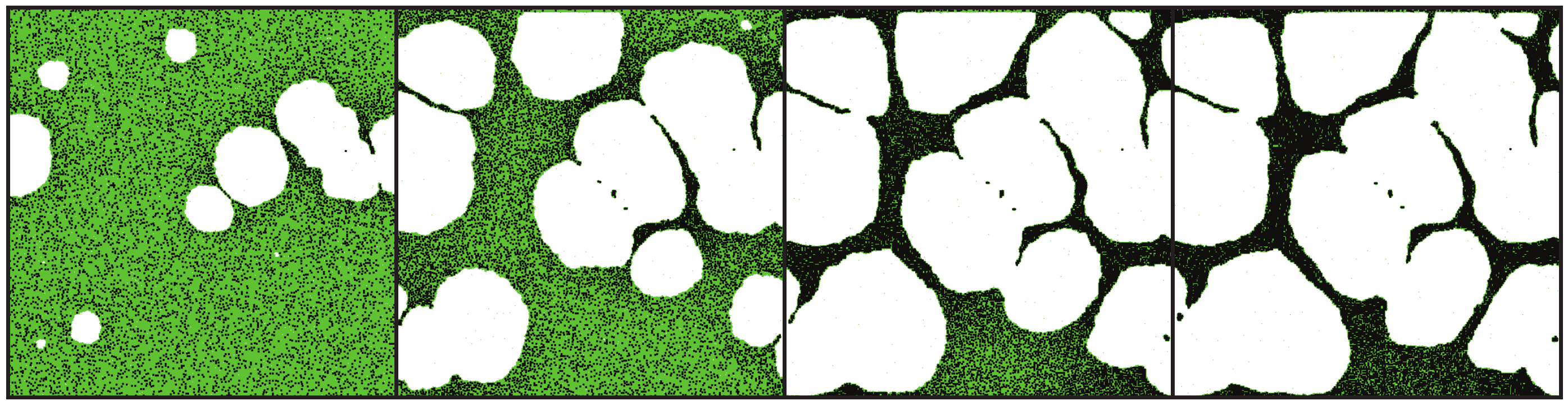}
\caption{
  Snapshots from typical evaporative dewetting processes with
  nanoparticles in the case of (a) a spinodal-like process at
  $\mu=-2.55$ and (b) of nucleation and growth of holes at $\mu=-2.3$.
  Starting condition homogeneous liquid film with homogeneously
  distributed particles.  The number of Monte Carlo steps is indicated
  in the individual panels (top: 10, 15, 20, 100 bottom: 100, 200,
  300, 400); $\varepsilon_{nn}=2$, $\varepsilon_{nl}=1.5$, $M=50$,
  $\phi=0.2$.  The remaining parameters are as in
  Fig.~\ref{fig:withoutnanopart}.  Particles are black, liquid is gray
  (green online) and the empty substrate is white.  }
\mylab{fig:withnanopart}
\end{figure}

The aspect that most interests us here is, however, not the equilibrium
behavior but the dynamics of the liquid-gas phase transition. As in
Section~\ref{sec:won} we start at a $\mu<-2$ with a liquid covered
substrate, i.e., in a region where the final state without
nanoparticles corresponds to the vapor phase. Now, however, although
the liquid evaporates the nanoparticles remain.  Depending on the
particular parameter values chosen one finds final structures that do
not change any more on short time scales. They might, however, slowly
evolve, e.g.~ by coarsening on long time scales as conditions are
'fluxional' \cite{RRGB03}. The final 'dried-in' structures depend on
the pathway of evaporative dewetting (i.e.~on $kT$, $\mu$, the
interaction constants $\varepsilon_{ij}$, nanoparticle concentration
$\phi$ and mobility $M$) and range from labyrinthine to polygonal
network structures or holes in a dense particle layer. Typical
snapshots from the evolution of a layer of solution are shown in
Fig.~\ref{fig:withnanopart}.  As before, the evaporative dewetting
process of the solvent follows either a nucleation and growth
[Fig.~\ref{fig:withnanopart}(b)] or a spinodal-like
[Fig.~\ref{fig:withnanopart}(a)] pathway. Following the mean field
argument as above we expect for the limiting curve $\mu_{\rm ms}(kT)$
separating the two processes in the parameter plane again a shift by
$-2(\varepsilon_{nl}-1)\phi$. However, due to fluctuations this border
is elusive when scrutinized numerically, therefore we will not try to
compare simulations to the prediction.

At first sight, however, one might get the impression that the
particles act as a type of passive tracer that preserves the transient
volatile dewetting structures of the solvent. This idea was put
forward in Refs.~\cite{Mert97,Mert98,TMP98} in the context of
experiments on dewetting aqueous solutions of macromolecules and
provides an explanation for some of the basic features of the observed
network structures. One can also employ this hypothesis to explain
some of the structures observed in the present study, such as the
network and spinodal structures shown in
Fig.~\ref{fig:withnanopart}. The simulations indicate, however, that
the nanoparticles are not simply passive tracers.  Although the
particles primarily just follow the solvent, they play an important
role in several phases of the process.

\begin{figure}[htbp]
\centering
\includegraphics[width=0.7\hsize]{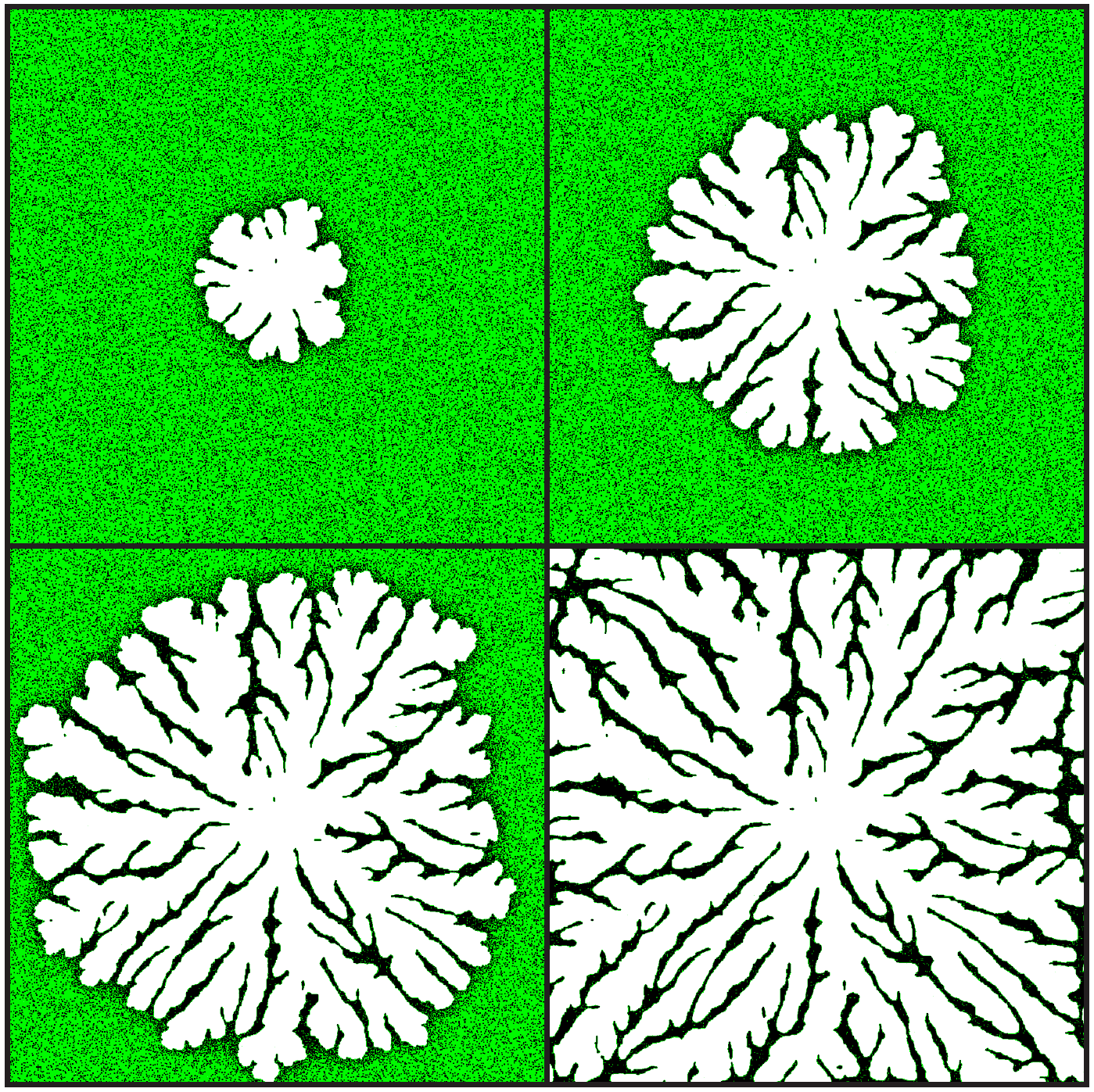}\hspace{.2cm}
\caption{Snapshots illustrating the unstable growth of a nucleated
  hole in a film of nanoparticle solution. The strong branched
  fingering caused by the nanoparticles can be seen clearly. From top
  left to bottom right the number of Monte Carlo steps is 450, 1200,
  1800, and 3000. Nanoparticle related parameters are
  $\varepsilon_{nn}=2$, $\varepsilon_{nl}=1.5$, $\mu=-2.3$, $M=5$,
  $\phi=0.2$.  The remaining parameters are as in
  Fig.~\ref{fig:withoutnanopart}.  Particles are black, liquid is gray
  (green online) and the empty substrate is white.}
\mylab{fig:fingering}
\end{figure}

First, the particles may influence the nucleation process in the
metastable parameter region, beyond the 'shift' in $\mu$ predicted by
the mean field considerations above. Although our simulations seem to
indicate that the nucleation rates actually depend on the particles
themselves, we cannot quantify the effect at present. We leave this
question for future investigations and focus here on the second and
more pronounced influence.
Second, for a low mobility of the particles, i.e., for a slow
diffusion of particles as compared to the evaporation of the solvent,
the nucleated holes might grow in an unstable manner as illustrated in
Fig.~\ref{fig:fingering}. The transverse instability of
the dewetting front responsible for this effect is analyzed in more detail in the following
section.
%
%%%%%%%%%%%%%%%%%%%%%%%%%%%%%%%%%%%%%%%%%%%%%%%%%%%%%%%%%%%%%%%%%%%%%%%
\section{The fingering instability}
\mylab{sec:finger}
%%%%%%%%%%%%%%%%%%%%%%%%%%%%%%%%%%%%%%%%%%%%%%%%%%%%%%%%%%%%%%%%%%%%%%%%
%
A straight evaporative dewetting front without nanoparticles present
moves with a constant mean velocity $\bar{v}$ (averaged over the
transverse direction). The velocity only depends on the driving
chemical potential. The local velocity, however, fluctuates with temperature. 
The fluctuations are illustrated in the column of snapshots at the right of Fig.~\ref{fig:phase}.  Therefore,
the local front position also fluctuates around the steadily advancing or
receding mean position.  The velocity is easily measured numerically
for driving forces small enough that the nucleation of additional
holes is not very probable during the time of measurement. Here this
is the case for $\mu \ge -2.4$. The dependency of $\bar{v}$ on $\mu$ is
given for comparison below in Fig.~\ref{fig:velocity}. One finds that
the velocity increases linearly with the driving force.

Adding diffusing nanoparticles to the solvent changes the front
behavior dramatically as it may become transversally unstable. Such
an unstable front does not only fluctuate around a continuously
receding mean position but fingers stay behind permanently (for the
front instability of a growing circular hole see Fig.~\ref{fig:fingering} above).
To elucidate the underlying mechanism we will in the next paragraphs discuss the influence of the individual
system parameters on the instability, i.e., the influence of chemical
potential $\mu$, particle concentration $\phi$, and mobility $M$.

For each dependency we present snapshots of the final dried in
nanoparticle structure corresponding to the dried in patterns observed
in experiment. We add, however, thin lines that correspond to
positions of the dewetting front at equidistant times throughout the
evolution. The final branched patterns we characterize in a rather
simple but effective manner. We determine an averaged finger
number by dividing the side length of the computational square
domain by the average finger number.  The latter is obtained by
counting the fingers on each line orthogonal to the mean direction of
front motion (here the $y$ direction) and averaging over the $y$-range
where fingers exist. The underlying hypothesis that the finger
patterns are 'stationary' will be discussed and checked below. Attempts
to determine a mean distance of fingers employing a 1-D Fourier
transform unfortunately did not give meaningful results because of the strong
side-branching (see snapshots below).  Note, that the finger number
corresponds to a wave number rescaled by $2\pi/L$ where $L$ is the
side length of the computational domain (that is fixed throughout this
section and corresponds to about $1.2\mu$m.).
\subsection{Dependence on chemical potential}
\mylab{sec:fing-mu}

\begin{figure}[htbp]
(a)\includegraphics[width=0.4\hsize]{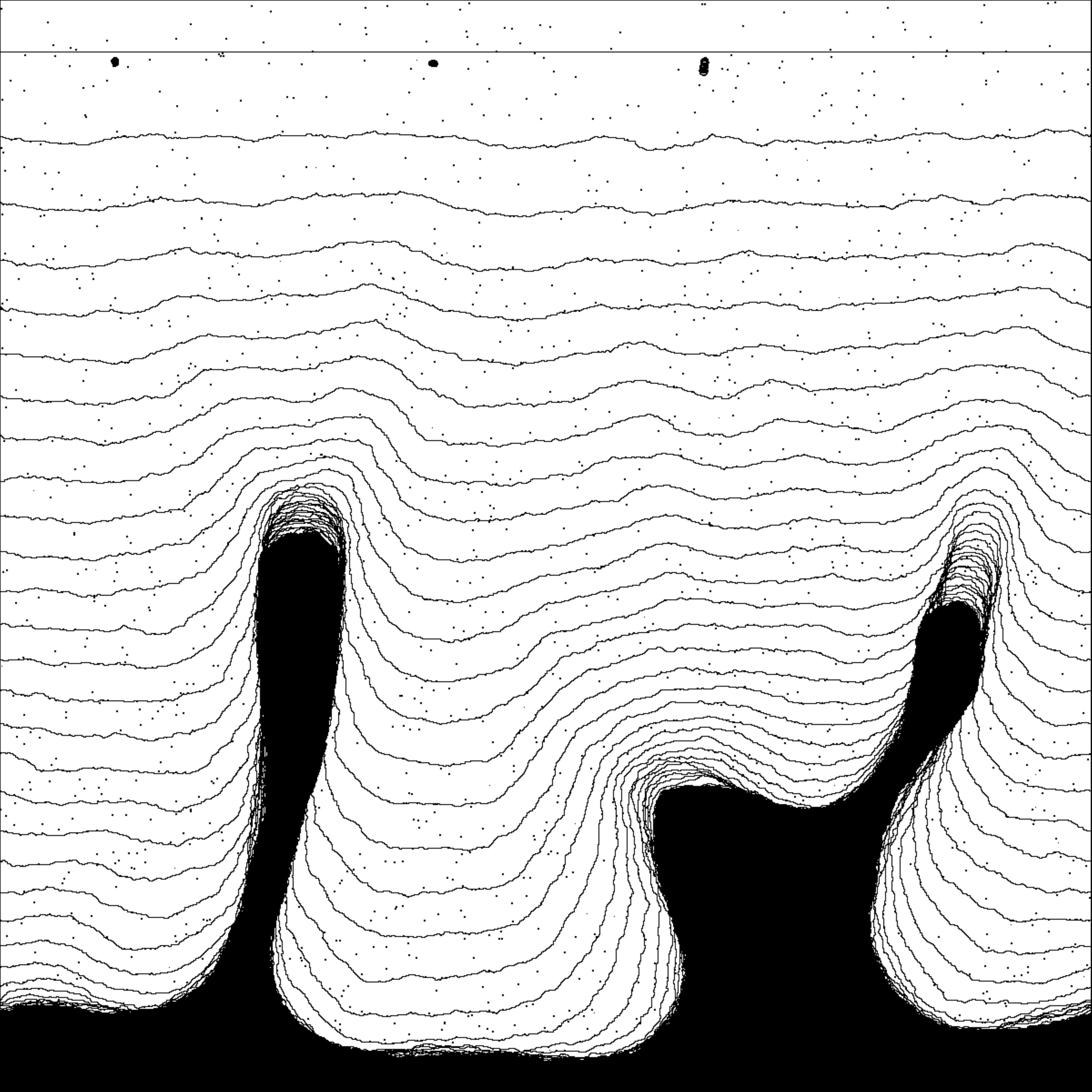}\hspace{0.025\hsize}
\includegraphics[width=0.4\hsize]{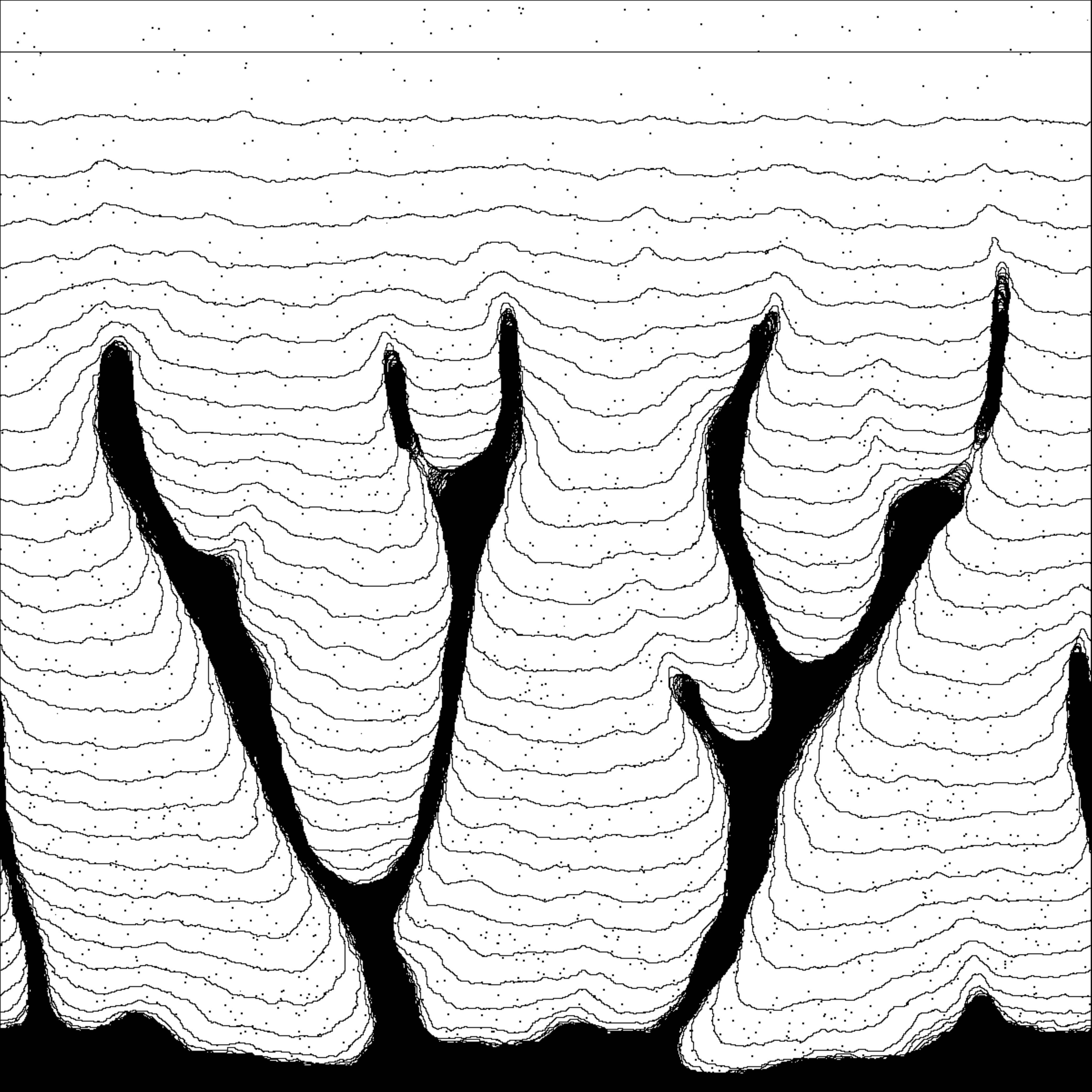}(b)\\
(c)\includegraphics[width=0.4\hsize]{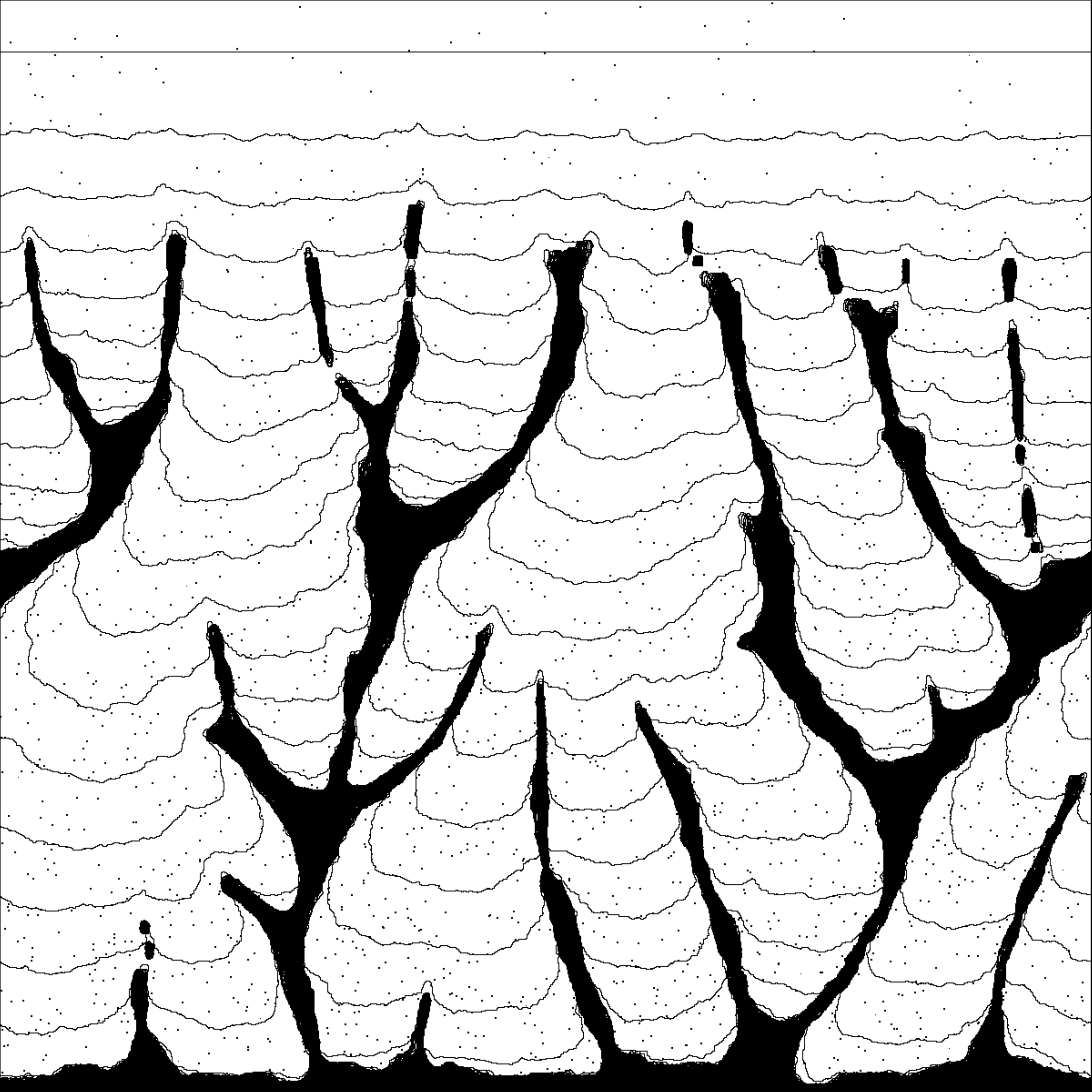}\hspace{0.025\hsize}
\includegraphics[width=0.4\hsize]{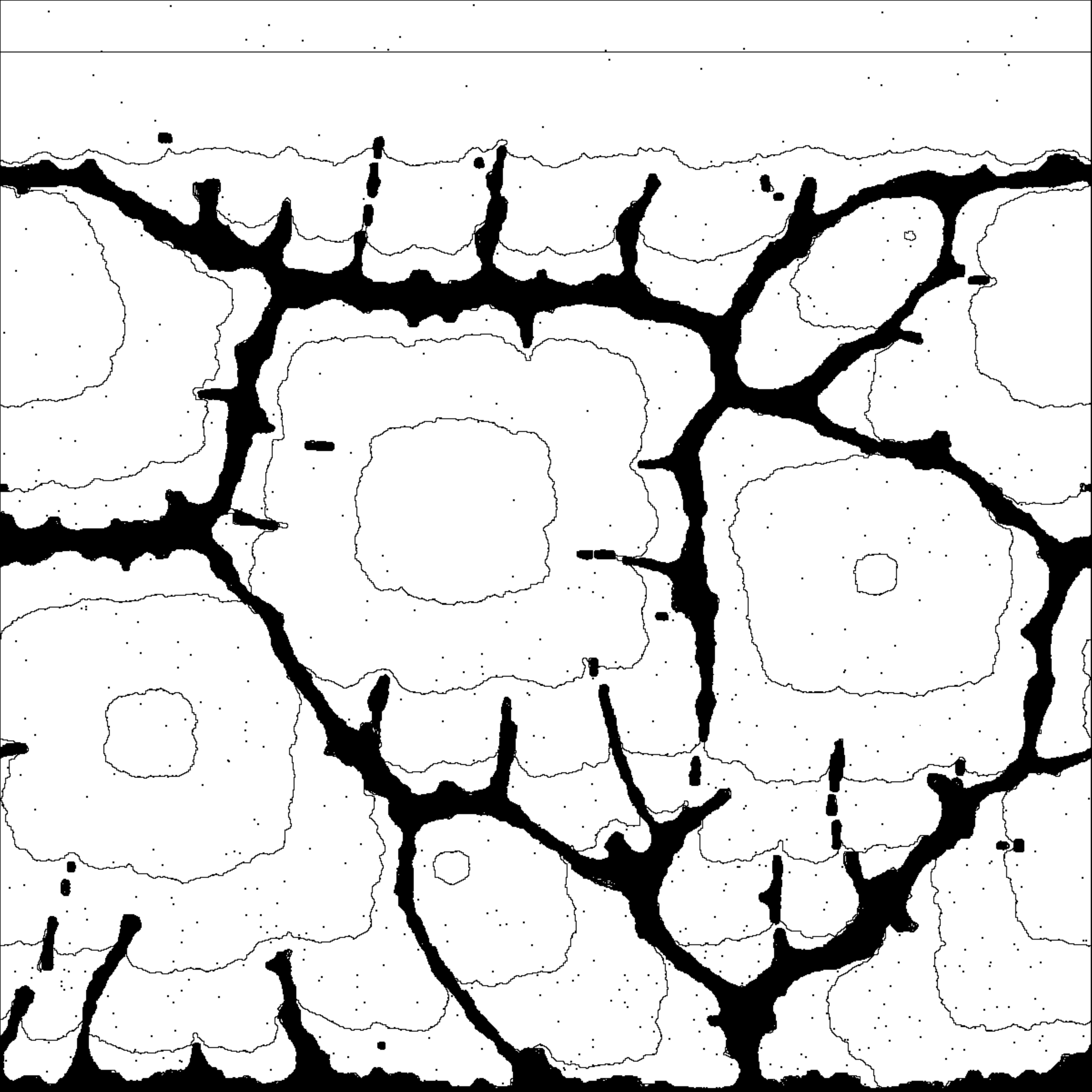}(d)
\caption{
  Final 'dried-in' branched fingering structure for evaporative
  dewetting of a nanoparticle solution for different driving forces,
  i.e., chemical potential (a) $\mu=-2.1$, (b) $\mu=-2.2$, (c)
  $\mu=-2.3$, and (d) $\mu=-2.4$.  Thin lines correspond to positions
  of the dewetting front at equidistant times with $\Delta t=333$ (a), and $\Delta t=166$ (b, c, d).
  The domain size is $1200\times1200$, $M=20$, $\phi=0.1$, $kT=0.2$,
  $\varepsilon_{nn}=2$, and $\varepsilon_{nl}=1.5$.}
\mylab{fig:contours-mu}
\end{figure}

\begin{figure}[htbp]
\includegraphics[width=0.8\hsize]{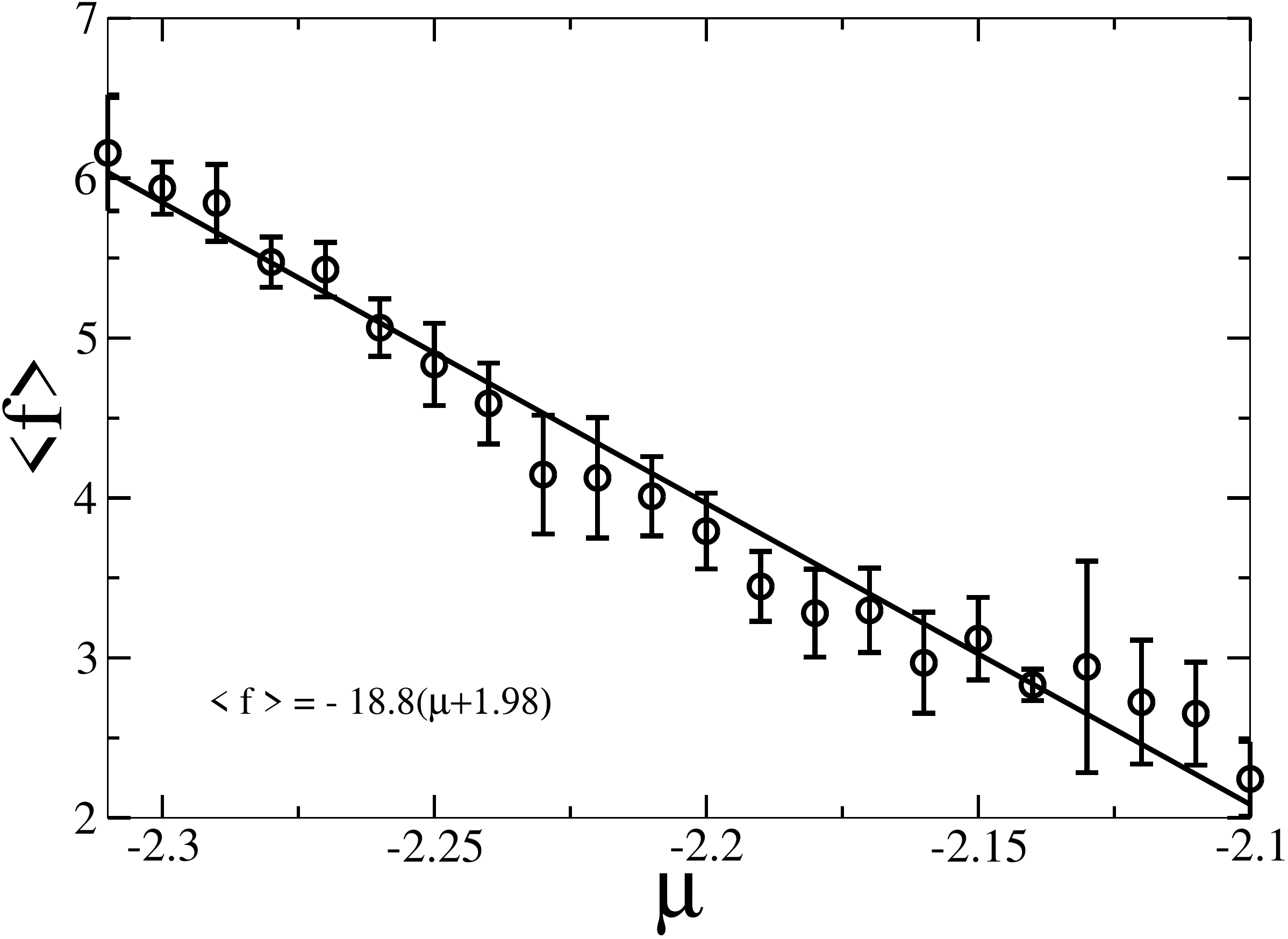}
\caption{Dependence of mean finger number $<f>$ of the final dried in structures
(determined as described in the main text) on chemical potential $\mu$.
Every data point corresponds to the mean value of 7 simulations; error
bars indicate the corresponding standard deviation.
The solid line corresponds to a linear mean-square fit with coefficients given in the plot.
The remaining parameters are as in Fig.~\ref{fig:contours-mu}.
}
\mylab{fig:fingers-mu}
\end{figure}

First, we focus on the influence of the driving force, i.e., the
chemical potential.  Fixing the particle concentration to $\phi=0.1$ and the 
mobility to $M=20$, the final structures observed for various values of $\mu$ are shown in
Fig.~\ref{fig:contours-mu} whereas Fig.~\ref{fig:fingers-mu} shows the dependence of the
mean finger number on the chemical potential. The
driving force increases for decreasing $\mu$ and the number of
fingers increases roughly linearly. Again, for $\mu\le-2.4$ nucleation becomes more
probable and we observe a random polygonal network.
The side-branches of the polygonal pattern clearly result
from the transverse instability [see Fig.~\ref{fig:contours-mu}(d)].

Further inspection of Fig.~\ref{fig:contours-mu} results in the
following observations:\\
(i) Both the wavenumber and the growth rate of the front
instability increase with decreasing $\mu$.  A faster growth
translates into longer fingers in the 'dried in' structure, i.e.,
fingers that extend further in Fig.~\ref{fig:contours-mu}. Starting
from the same initial front position, the fingers appear much earlier
in panel (c) than in panels (b) or (a).\\
(ii) The front contours taken at equidistant times indicate that the
front is already strongly disturbed before the instability becomes
manifest in a deposited finger, i.e., parts of the front slow down
before actually stopping (and depositing material). This is especially
well visible in Fig.~\ref{fig:contours-mu}(a).\\
(iii) The dried-in fingering structure is not entirely frozen as the
chosen parameters values are in a 'fluxional' regime
\cite{RRGB03}. This is especially well visible at the tips of the
fingers in Fig.~\ref{fig:contours-mu}(a) that continue to retract very
slowly, or in Fig.~\ref{fig:contours-mu}(b) where a thin part of the
rightmost finger breaks and slowly retracts as well. Eventually, this
will lead to a long-time coarsening of the finger structure.\\
(iv) It is remarkable that even when fingers are left behind the local
distance between consecutive front contours seems to be constant in
each of the panels [beside the regions discussed in (iii)].  One
expects, however, that the front velocity depends on the local
particle concentration and the mobility of the particles. The
observation of a constant velocity indicates that the front
instability is related to an auto-optimization of the front
velocity. This is analogous to an effect described for dewetting
polymer films \cite{ReSh01}. There, a moving dewetting front 'expels'
part of the liquid rim and leaves the liquid behind in the form of
deposited droplets. The instability avoids a slowing down of the
dewetting front.  Here, some of the particles that the evaporative
dewetting front collects are expelled and deposited in fingers when too
many are collected by the front.

Measurements of the front velocity for different $\mu$ and particle
concentrations indeed show that the mean velocity is constant and
depends linearly on $\mu$ (see Fig.~\ref{fig:velocity} below).

\subsection{Dependence on particle concentration}
\mylab{sec:fing-conc}

\begin{figure}[htbp]
(a)\includegraphics[width=0.4\hsize]{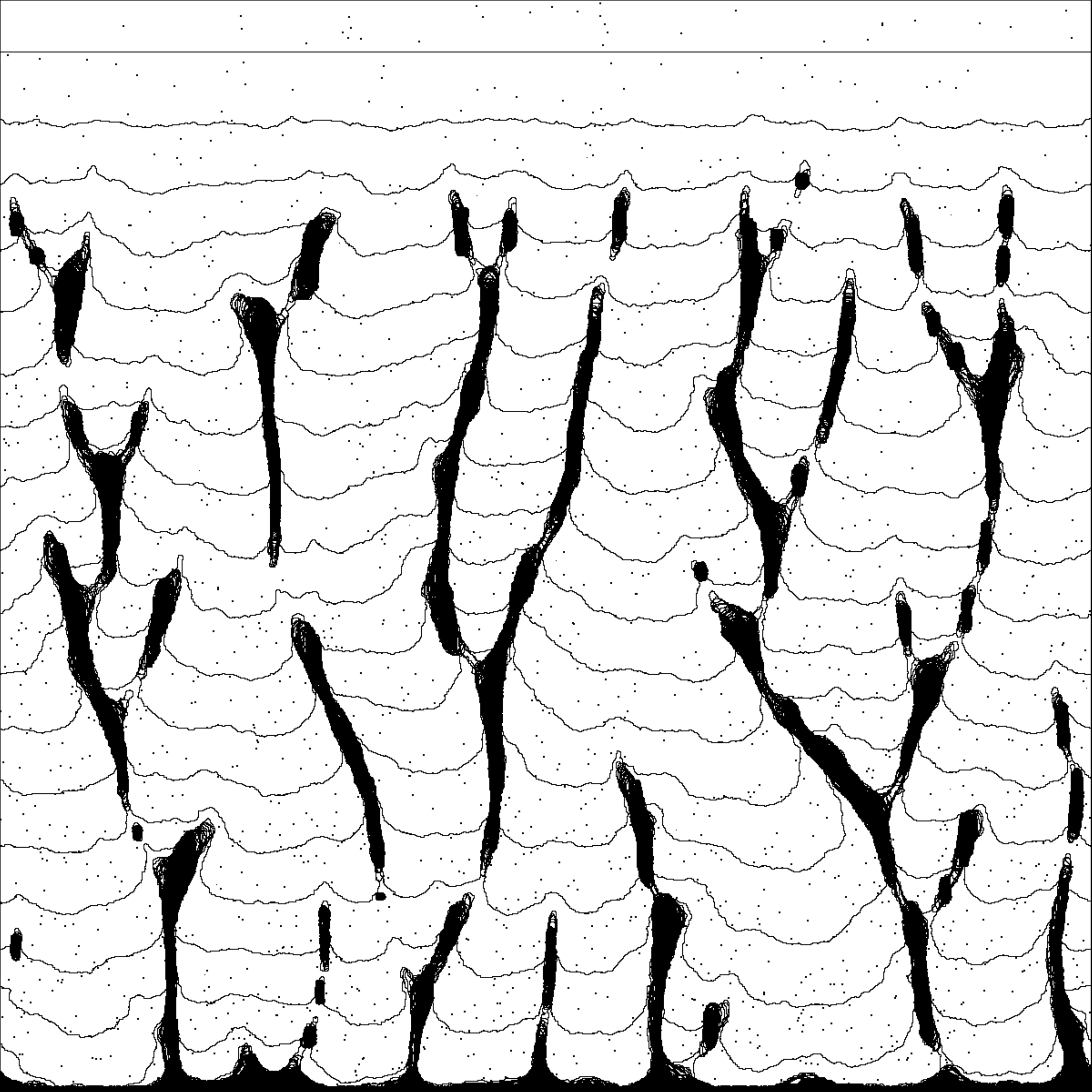}\hspace{0.025\hsize}
\includegraphics[width=0.4\hsize]{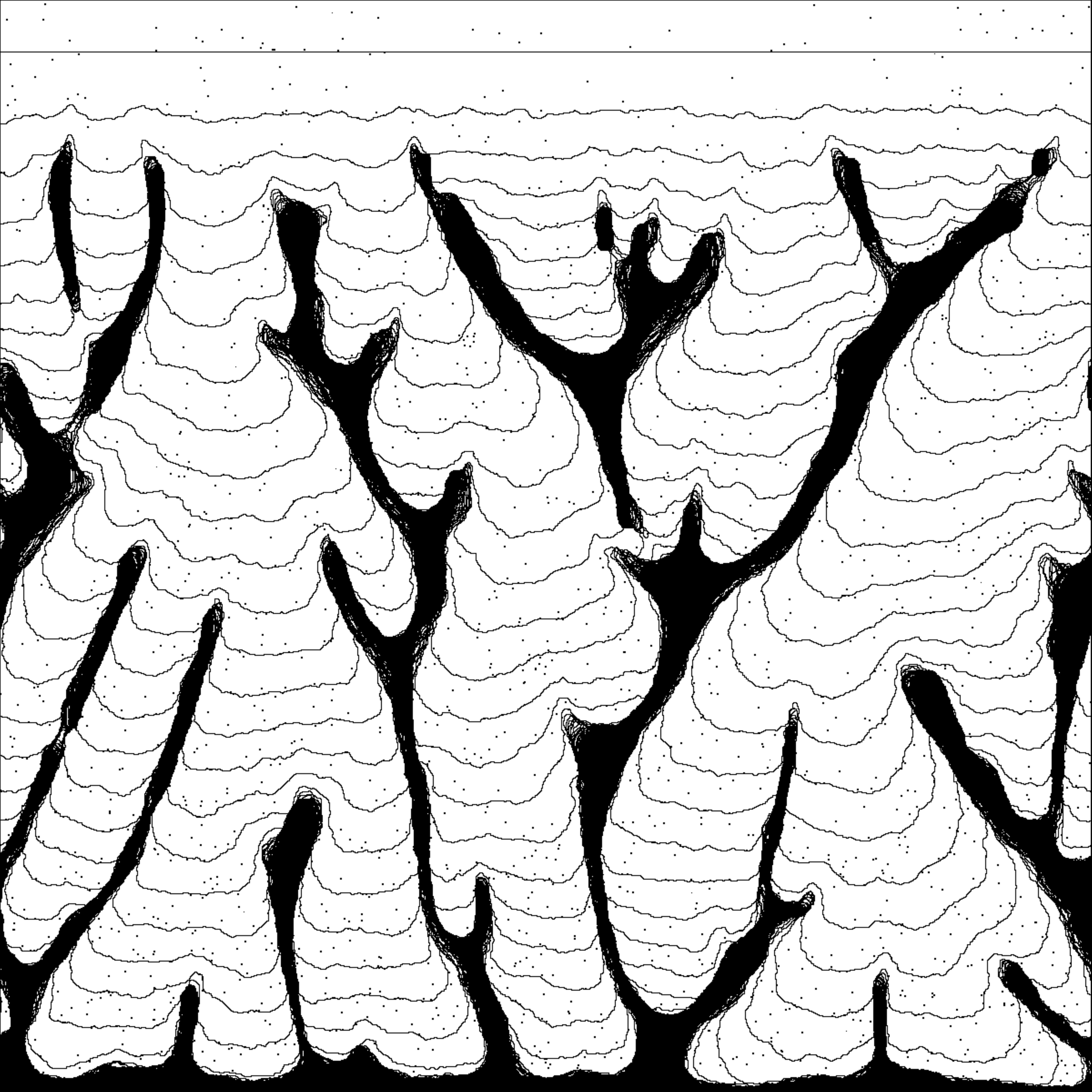}(b)\\
(c)\includegraphics[width=0.4\hsize]{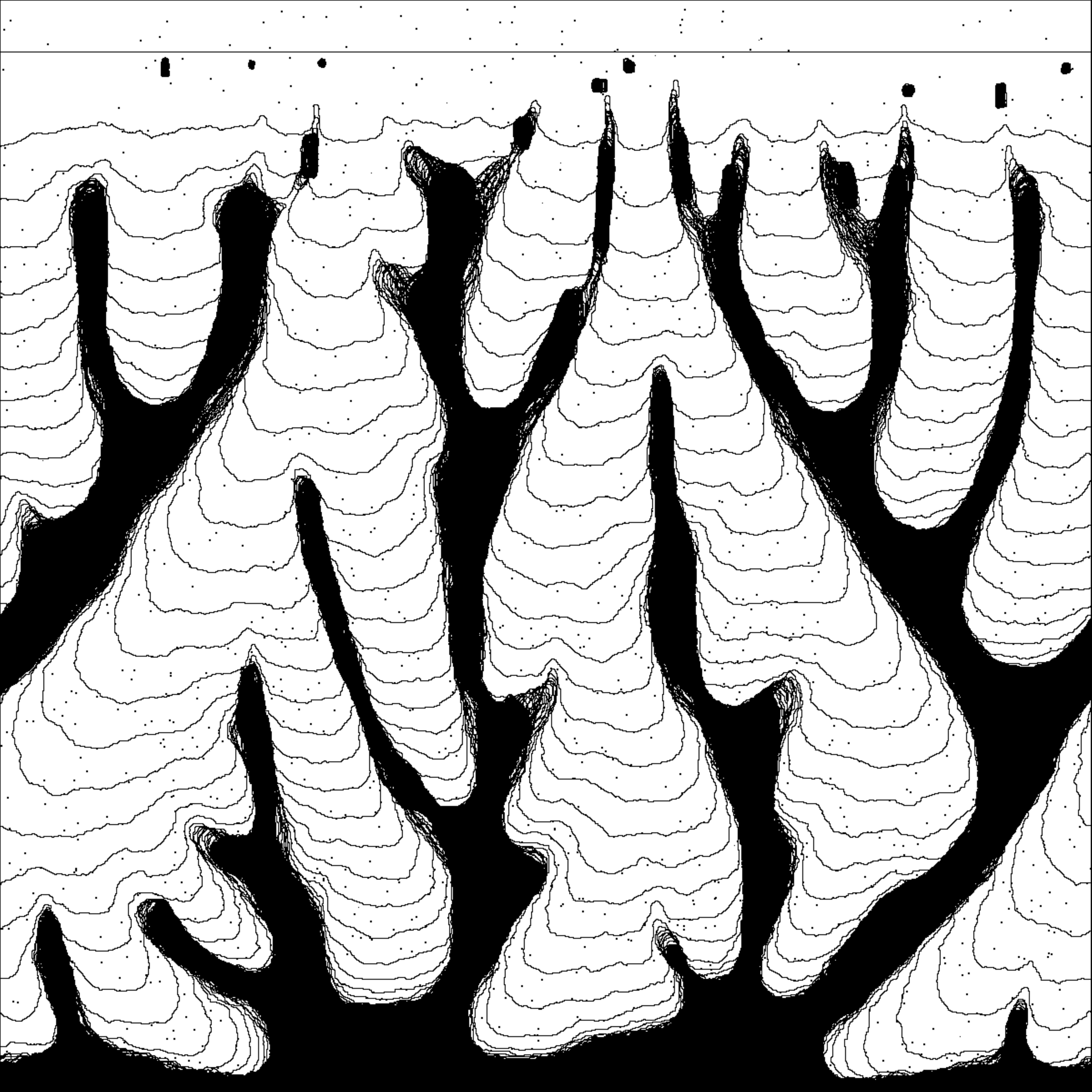}\hspace{0.025\hsize}
\includegraphics[width=0.4\hsize]{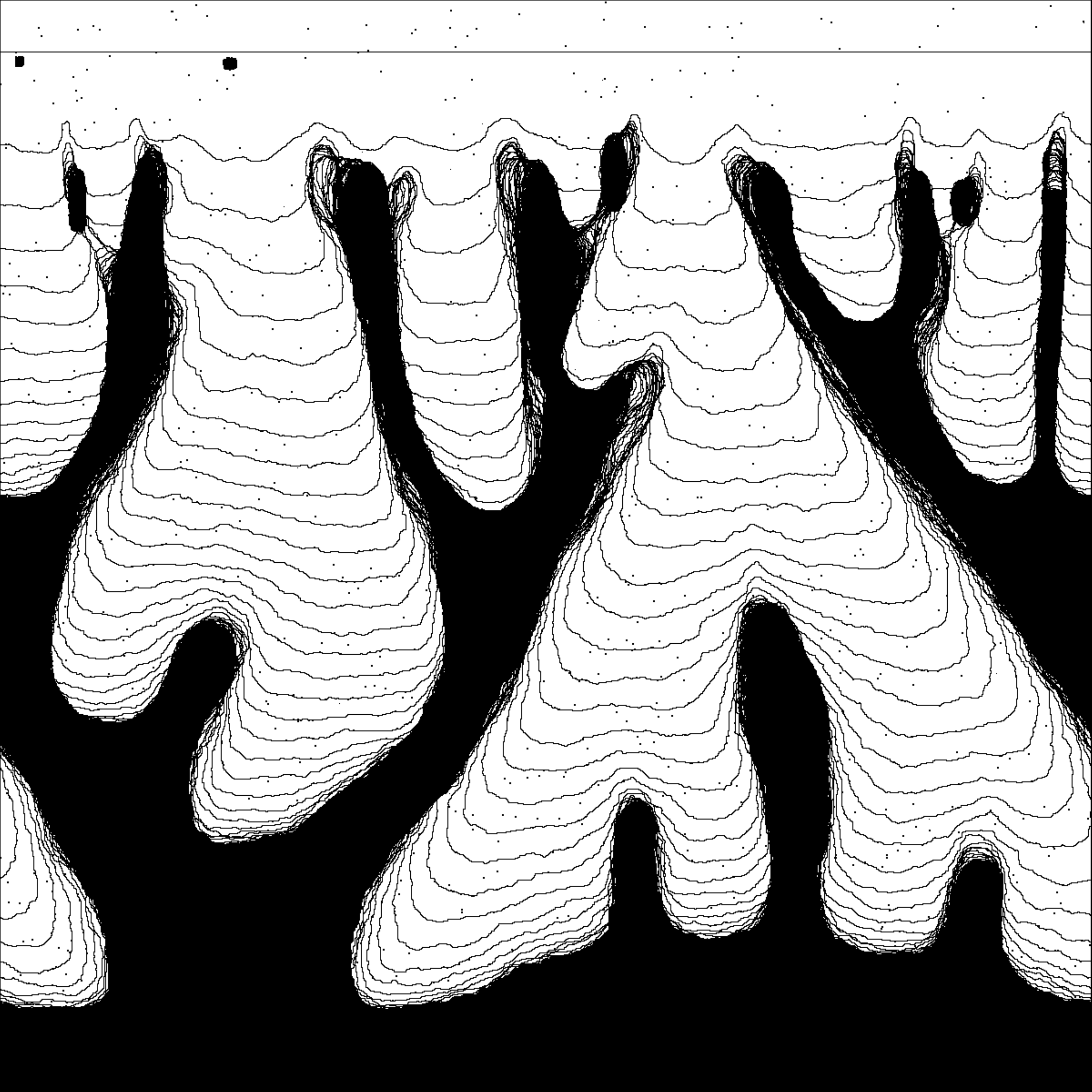}(d)
\caption{
  Final 'dried in' branched fingering structure for evaporative
  dewetting of a nanoparticle solution for different nanoparticle concentration
  (a) $\phi=0.05$, (b) $\phi=0.10$, (c) $\phi=0.20$,
and (d) $\phi=0.30$.  Thin lines correspond to positions
  of the dewetting front at equidistant times with $\Delta t=166$ (a, b), $\Delta t=333$ (c) and $\Delta t=500$ (d).
  The domain size is $1200\times1200$, $M=10$, $\mu=-2.2$, $kT=0.2$,
  $\varepsilon_{nn}=2$, and $\varepsilon_{nl}=1.5$.}
\mylab{fig:contours-conc}
\end{figure}

We have also examined the influence of the nanoparticle concentration
$\phi$ on the fingering.  Fixing the chemical potential at $\mu=-2.2$ and the
mobility at $M=10$, the final structures are shown for various values of $\phi$ in
Fig.~\ref{fig:contours-conc}.
From an initial inspection of Fig.~\ref{fig:contours-conc} one might get the
erroneous impression that the finger number increases with
concentration. However, the statistics of many runs shows that the
number is nearly independent of the particle concentration. The
fingers only become thicker with increasing concentration. This
indicates that the fingering is controlled by the 'dynamical'
parameters responsible for the ratio of the time scales for diffusion
of the particles and motion of the evaporative front (i.e., mobility
and chemical potential).  As the influence of the chemical potential
has already been established in Section~\ref{sec:fing-mu} we focus
below in Section~\ref{sec:fing-mob} on the influence of mobility.

\begin{figure}[htbp]
\includegraphics[width=0.8\hsize]{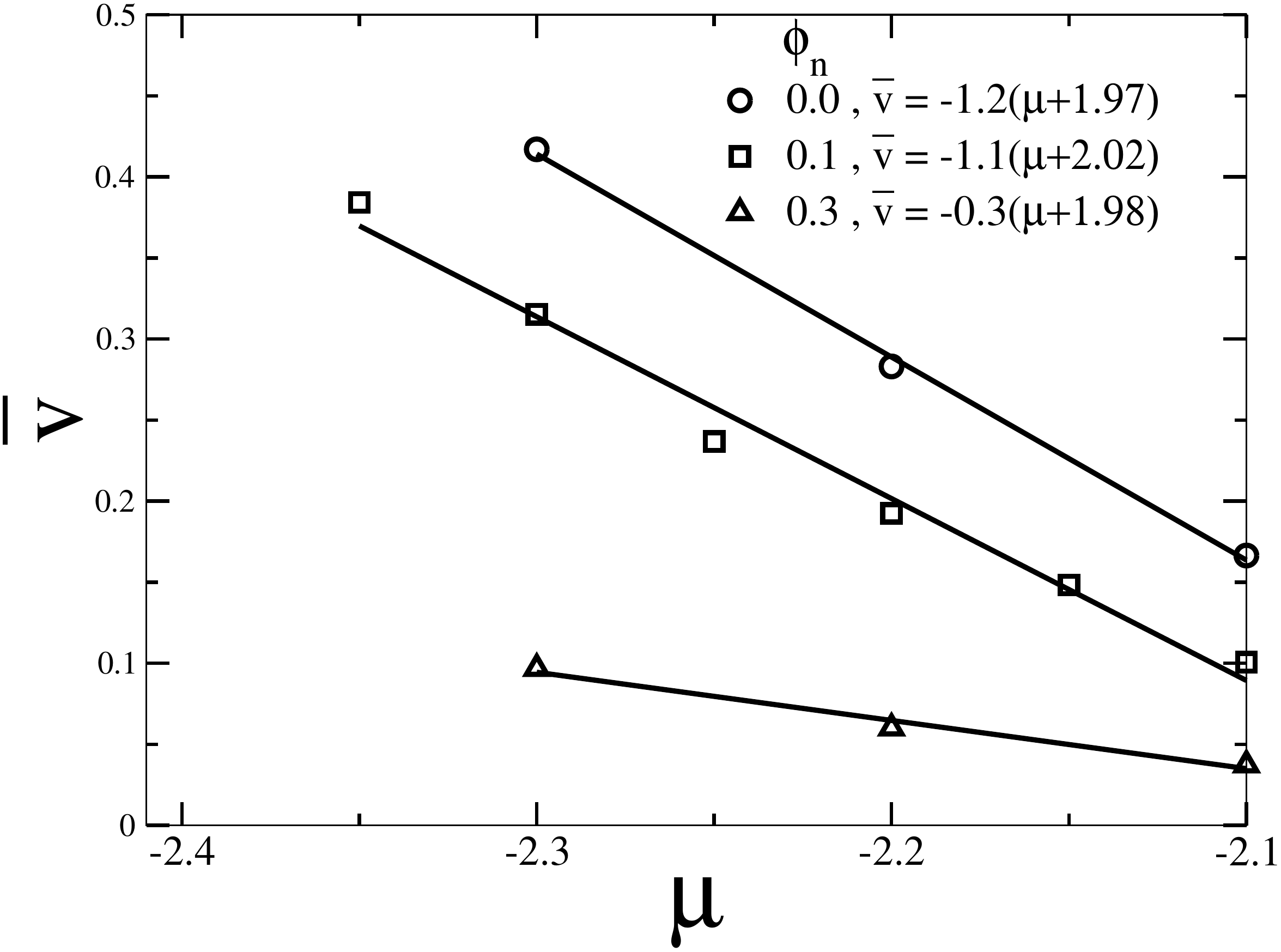}
\caption{
  Dependence of mean front velocity on chemical potential $\mu$ and
  for selected particle concentrations $\phi$ as given in the
  legend. The thin lines correspond to linear regression fits with coefficients
  as given in the plot. The used domain size is $1200\times1200$,
  $M=20$, $kT=0.2$, $\varepsilon_{nn}=2$, and
  $\varepsilon_{nl}=1.5$. The velocity is averaged over 7000 MC steps
  or the number of steps it takes to 'dry ' the substrate, whichever is
  smaller.}
\mylab{fig:velocity}
\end{figure}

Before we do so we summarize in Fig.~\ref{fig:velocity} results for
the dependence of the global average front velocity on chemical
potential and particle concentration.  For all investigated
concentrations the velocity decreases linearly with increasing
chemical potential, i.e., with decreasing driving force. If we fix the
chemical potential, the dewetting process is slower for higher particle
concentrations. This implies that the finger number is not directly
related to the mean front velocity, a result that will be discussed
in the Conclusion.
\subsection{Dependence on mobility}
\mylab{sec:fing-mob}

\begin{figure}[htbp]
(a)\includegraphics[width=0.4\hsize]{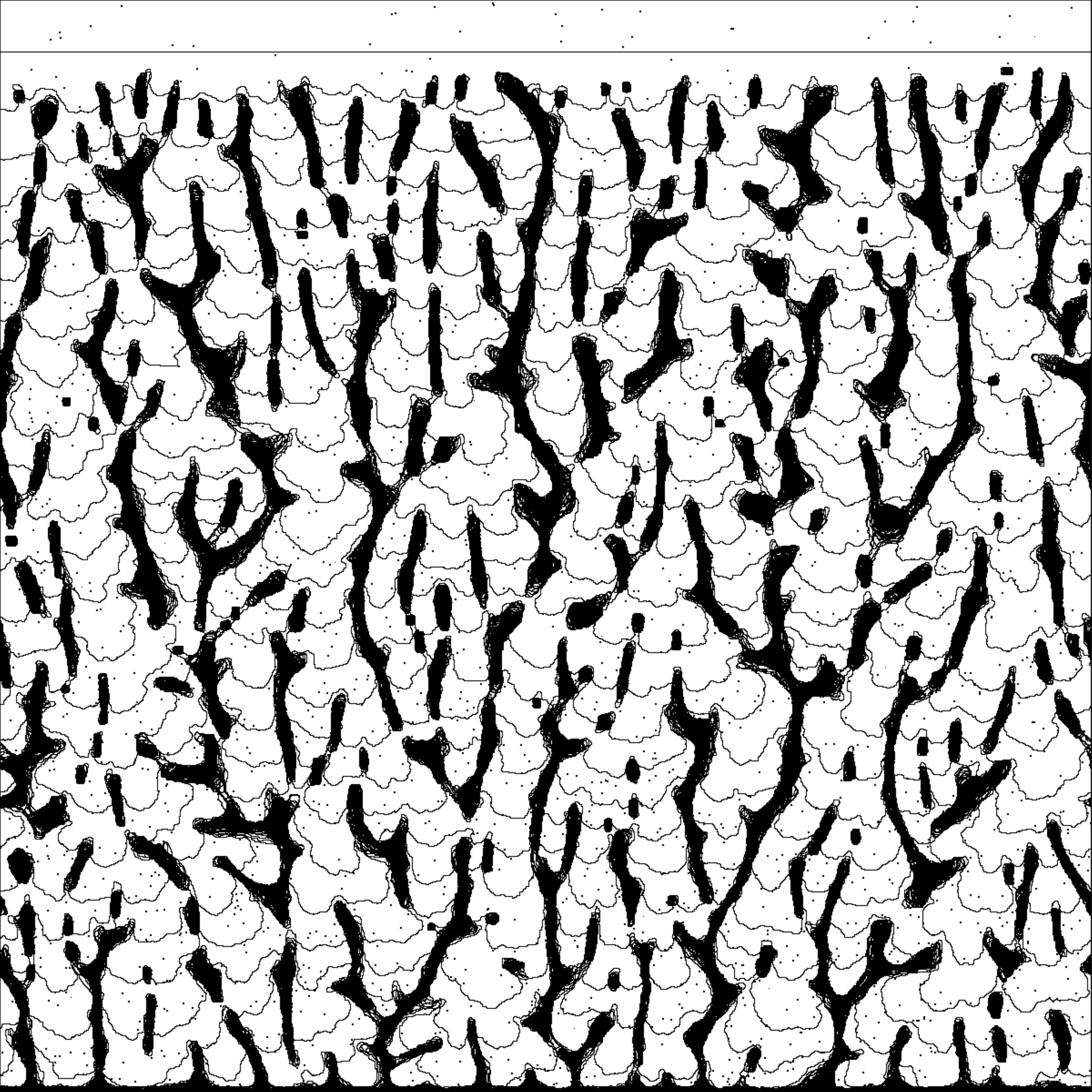}\hspace{0.025\hsize}
\includegraphics[width=0.4\hsize]{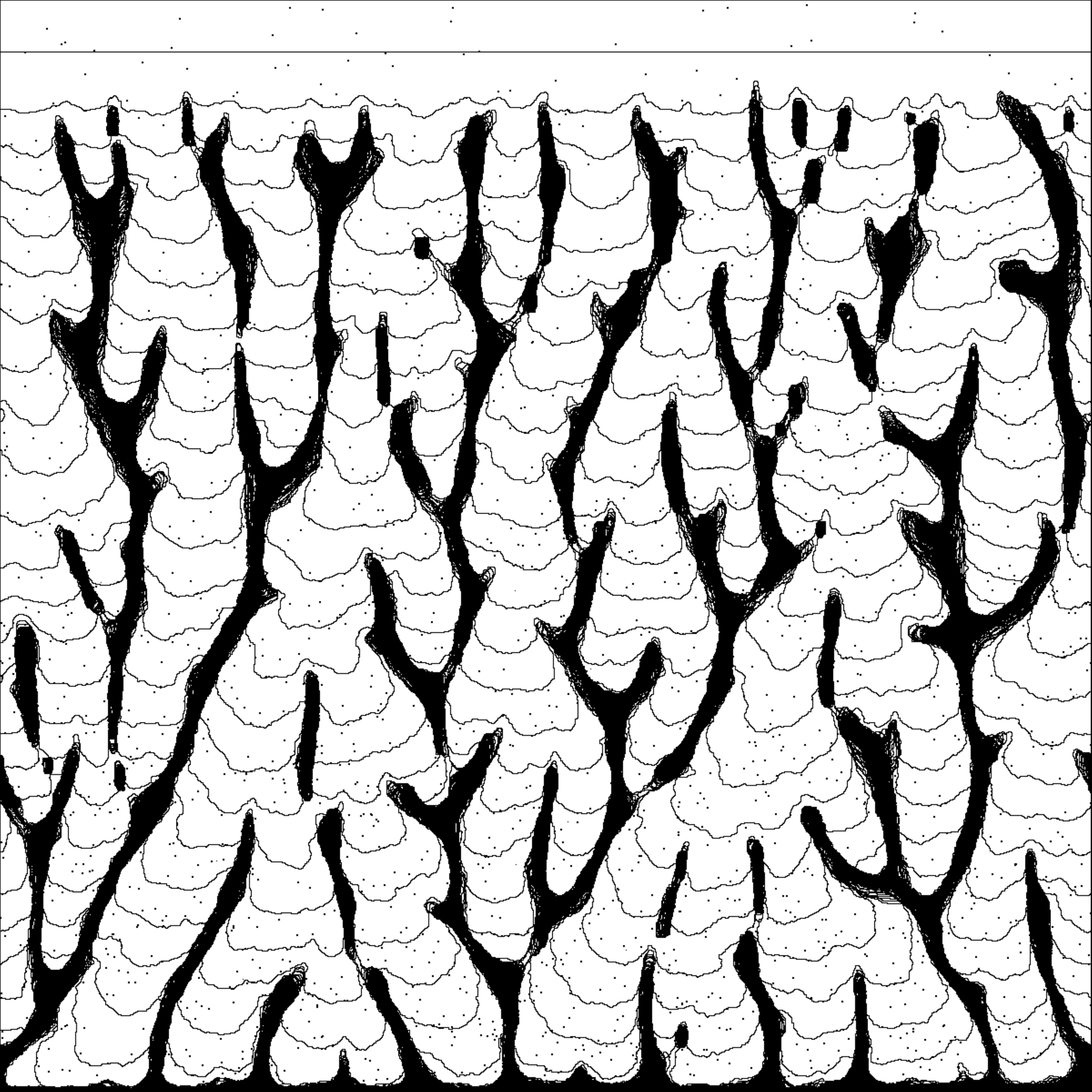}(b)\\
(b)\includegraphics[width=0.4\hsize]{k5000_M10_el1_en2_enl1,5_mu-2,2_cov0,1_kT0,2}\hspace{0.025\hsize}
\includegraphics[width=0.4\hsize]{k5000_M20_el1_en2_enl1,5_mu-2,2_cov0,1_kT0,2}(d)
\caption{Final dried in branched fingering structure for evaporative
  dewetting of a nanoparticle solution for different nanoparticle mobilities
  (a) $M=2$, (b) $M=5$, (c) $ M=10$, and (d) $M=20$.  Thin lines correspond to
  positions of the dewetting front at equidistant times with $\Delta t=166$.
  The domain size is $1200\times1200$, $\phi=0.1$, $\mu=-2.2$, $kT=0.2$,
  $\varepsilon_{nn}=2$, and $\varepsilon_{nl}=1.5$.}
\mylab{fig:contours-mob}
\end{figure}

\begin{figure}[htbp]
\includegraphics[width=0.8\hsize]{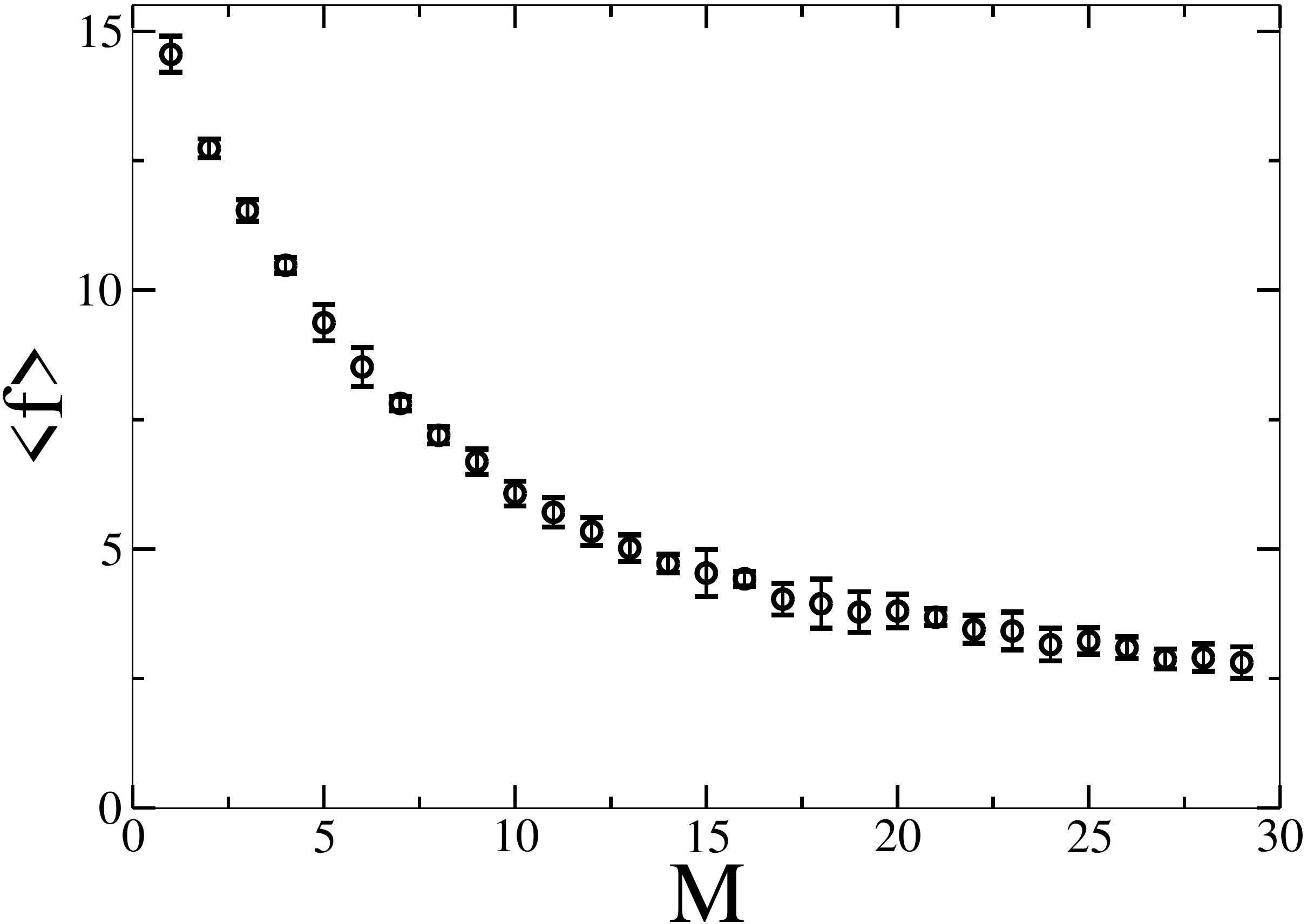}
\caption{Dependence of mean finger number $<f>$ of the final dried in
  structures on particle mobility $M$.  Every data point corresponds
  to the mean value of 7 simulations; error bars indicate the
  corresponding standard deviation. The remaining parameters are as in
  Fig.~\ref{fig:contours-mob}.}
\mylab{fig:fingers-mob}
\end{figure}

Fixing the particle concentration at $\phi=0.1$ and the chemical
potential at $\mu=-2.2$, we show in Fig.~\ref{fig:contours-mob} the 
final structures observed for selected values of mobility, $M$.
Fig.~\ref{fig:fingers-mob} shows the dependence of mean finger number on mobility.  One
sees at once that mobility is very influential in the fingering process. For
decreasing mobility of the particles the number of fingers increases
strongly. As was observed for the influence of the chemical
potential, the growth rate of the instability increases with
increasing finger number, i.e., decreasing mobility. This is manifest
in Fig.~\ref{fig:contours-mob} by the later emergence of fingers for
larger $M$, i.e., in the increasing distance of the finger tips from
the upper domain boundary.

The systematic picture that emerges from the preceding three sections
allows us to identify the primary factors influencing the fingering instability. It
becomes obvious that they are of dynamical nature.  Mobility is the
major player. The faster the nanoparticles can diffuse away from the
dewetting front, the less likely the instability becomes as the front
is not fast enough to 'collect' the particles efficiently. This directly
corresponds to the observation that the fingering becomes more intense
with increasing driving force, i.e., lower chemical potential. Then
the relative diffusivity is hold constant (mobility $M$) but the mean
front speed increases with decreasing $\mu$. The greater the number of particles
collected at the moving front, the more likely become lateral density
fluctuations that lead to self-amplifying variations of the
local front velocity.  Note, however, that up to this point we kept all the
interaction parameters $\varepsilon_{ij}$ fixed.  We expect them to
influence the feedback loop because, for example, a stronger particle-particle
interaction $\varepsilon_{nn}$ should support clustering. But one has to be cautious
because the process is of dynamic character. Therefore it is difficult to
predict whether a larger $\varepsilon_{nn}$ implies less or more
fingers. This question will be discussed below in Section~\ref{sec:demix}.

Next, however, we will check the assumption of stationarity of the
finger patterns that we have used above to introduce the measure of the
mean finger number.

%%%%%%%%%%%%%%%%%%%%%%%%%%%%%%%%%%%%%%%%%%%%%%%%%%%%%%%%%%%%%%%%%%%%%%%%
\subsection{Stationarity of finger pattern}
\mylab{sec:stat-fing}
%%%%%%%%%%%%%%%%%%%%%%%%%%%%%%%%%%%%%%%%%%%%%%%%%%%%%%%%%%%%%%%%%%%%%%%%

\begin{figure}[htbp]
\includegraphics[width=0.8\hsize]{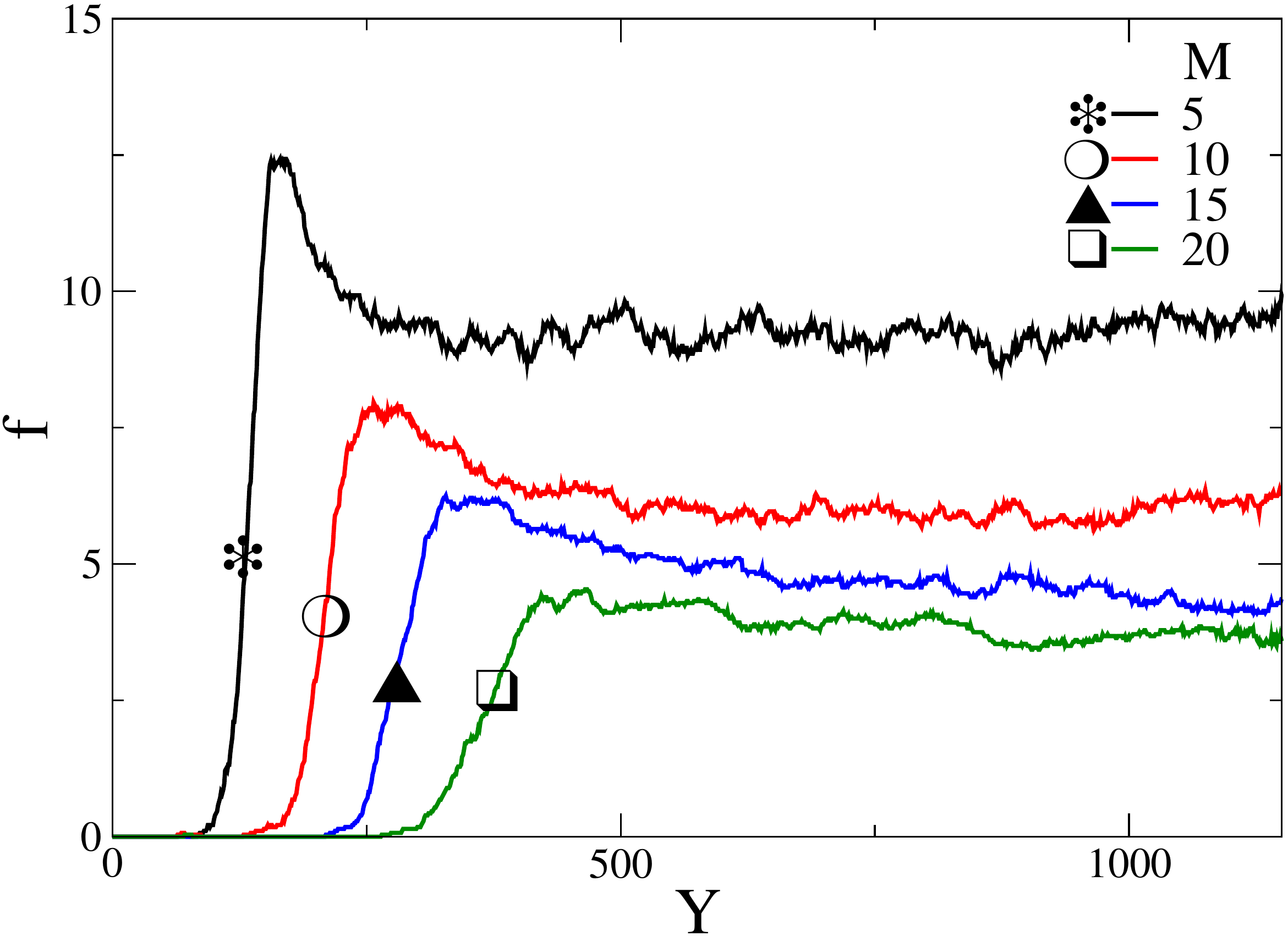}
\caption{Dependence of finger number on streamwise coordinate (i.e.,
  $y$-direction) for various values of mobility as given in the
  legend. As the structures are fixed when they stay behind the front
  increasing $y$ also indicates increasing time. The figure may
 be read as a space-time plot of the fingering at the moving
  dewetting front.  Each curve corresponds to the average of 28
  runs. The remaining parameters are as in Fig.~\ref{fig:contours-mob}.
}
\mylab{fig:fingers-time-mob}
\end{figure}

\begin{figure}[htbp]
\includegraphics[width=0.8\hsize]{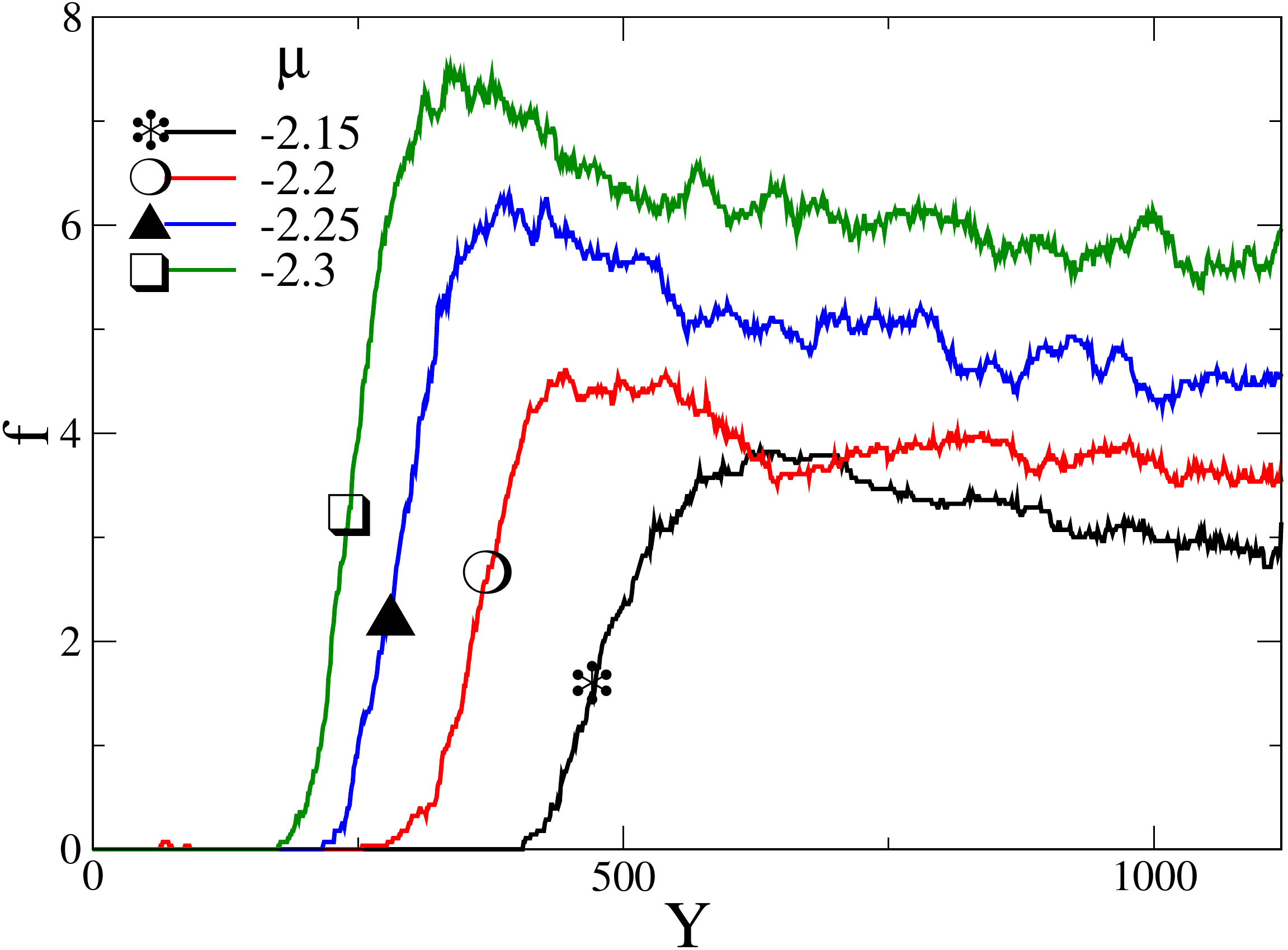}
\caption{Dependence of finger number on streamwise coordinate (i.e.,
  $y$-direction) for various values of the chemical potential as given in the
  legend. Each curve corresponds to the average of 28
  runs. The remaining parameters are as in Fig.~\ref{fig:contours-mu}.
}
\mylab{fig:fingers-time-mu}
\end{figure}

In Sections~\ref{sec:fing-mu} to~\ref{sec:fing-mob} we have used
the mean finger number to characterize the parameter dependency of the
fingering patterns. We noted that it only represents a trustworthy
measure if the fingering process is stationary. By ''stationary'' we
mean that the main properties of the pattern do not depend on the
streamwise position. In other words the moving dewetting front moves
on average with a constant velocity and deposits on average the same
amount of material in the same number of fingers.

To check this strong assumption we plot in
Figs.~\ref{fig:fingers-time-mob} and ~\ref{fig:fingers-time-mu} the
dependence of finger number on the streamwise co-ordinate
($y$-co-ordinate) for various values of the mobility and chemical
potential, respectively. To smooth out fluctuations each curve
represents the mean of 28 runs. As the fingers stay fixed after being
deposited behind the front, an advance in $y$ also indicates passing
time. We note in passing that all 'snapshots' above that
show dried-in final structures can as well be read as space-time plots
tracking the particle-deposition at the moving dewetting
front. Inspecting the curves we note that they have some properties
in common: (i) finger number starts from zero at some value $y=y_s$
well behind the initial position of the front at $y=60$; (ii) the
finger number grows exponentially on a typical length scale related to
the (linear) growth rate of the fingering instability; (iii) a maximal
finger number is reached before the number decreases again and settles
onto a stationary level. The decay corresponds to a small coarsening
of fingers and can be seen in several of the snapshots above. (iv)
Fluctuations around the stationary level are small, i.e., on average
as many new fingers are created (branch tips in the final image) as
vanish when fingers join pairwise (at the branching points).

Our conclusion is that the mean finger number as used above is a valid
measure because the $y$-range where the initial overshooting occurs is
relatively small, and also the overshoot itself is normally below 20\%
of the mean stationary value. The resulting error is small as compared
to the natural variance that we found between different runs.  Note,
however, that for several curves in Fig.~\ref{fig:fingers-time-mu} the
decay after the initial maximum is relatively slow and the stationary
value is not reached for our system size. This corresponds to a very
slow coarsening of fingers as the front recedes.

\subsection{Influence of interaction parameters}
\mylab{sec:demix}

We have mentioned in Section~\ref{sec:hom} the hypothesis (put forward in
\cite{TMP98} for a related system) that the ongoing pattern formation
and the properties of the receding dewetting front mainly result from
the dynamics of the liquid-gas phase transition. The nanoparticles
were seen as passive tracers. This describes well the situation in the
spinodal-regime (taking into account that the mean field parameters are
slightly changed by the particles). We have then shown that this
picture has to be amended because the particles have an important
influence during crucial phases of dewetting. Most notably, they
are solely responsible for the transverse instability of the dewetting front leading to the branched finger
patterns analyzed in detail in the present Section~\ref{sec:finger}.

The analysis up to now has been performed by fixing the interaction
parameters $\varepsilon_{ij}$ at convenient values: the liquid-liquid
interaction serves as an energy scale, i.e., $\varepsilon_{ll}=1$; the
particles attract each other slightly more than particles attract
liquid ($\varepsilon_{nn}=2$, $\varepsilon_{nl}=1.5$). This ensures
that particle-liquid phase separation does not generally occur in the
bulk solution, but also that particle clusters are likely to form at
higher particle concentrations. This resembles the situation in the
experiments and was also used in most computations in Refs.~\cite{RRGB03,MBM04}.

\begin{figure}[htbp]
\includegraphics[width=0.8\hsize]{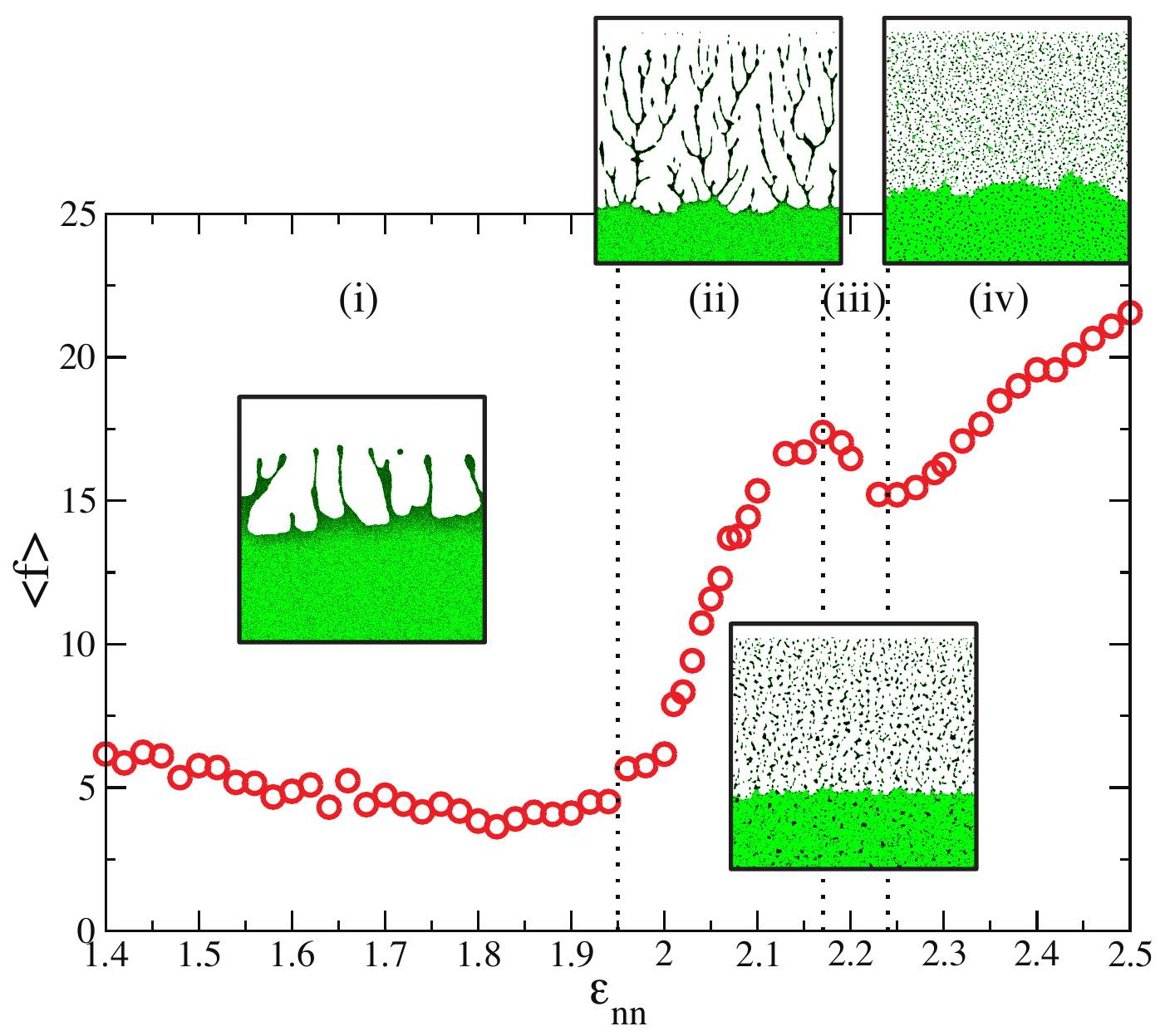}
\caption{Dependence of the mean finger number on the particle-particle
  interaction strength $\varepsilon_{nn}$. The regions marked (i) to
  (iv) are discussed in the main text.  The remaining parameters are:
  domain size $1200\times1200$, $kT=0.2$, $M=20$, $\mu=-2.2$,
  $\phi=0.1$, $\varepsilon_{nl}=1.5$. The insets give typical
  snapshots obtained in the four different regions. Particles are
  black, liquid is gray (green online) and the empty substrate is white.}
\mylab{fig:fingers-enn}
\end{figure}

\begin{figure}[htbp]
\includegraphics[width=0.8\hsize]{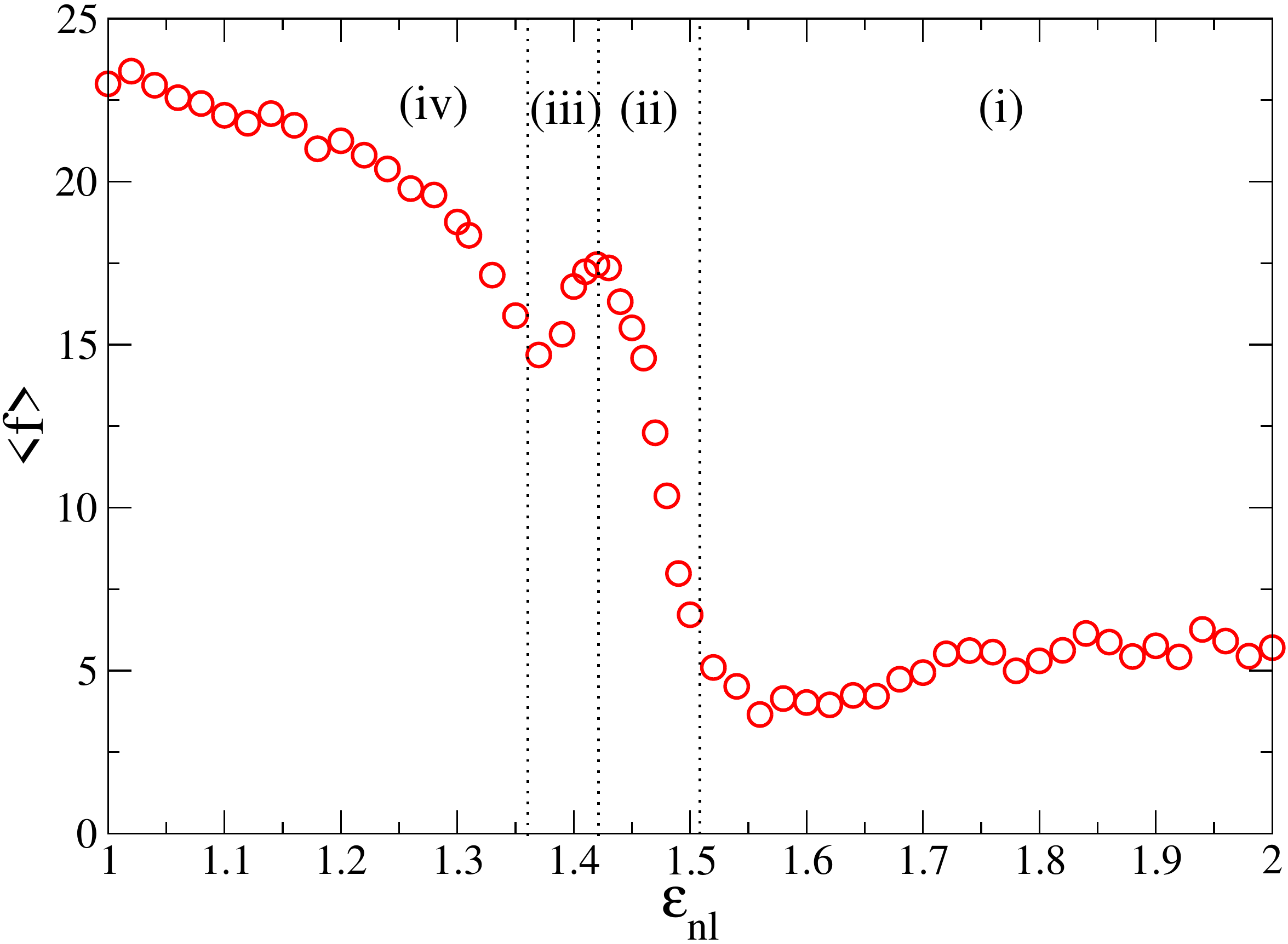}
\caption{Dependence of the mean finger number on the particle-liquid
  interaction strength $\varepsilon_{nl}$. The regions marked (i) to
  (iv) are discussed in the main text.  The remaining parameters are
  as in Fig.~\ref{fig:fingers-enn} and $\varepsilon_{nn}=2.0$.}
\mylab{fig:fingers-enl}
\end{figure}

In the following we will Analise the influence of $\varepsilon_{nn}$ and $\varepsilon_{nl}$ on the fingering
instability. In particular, we are interested in a possible coupling
of the front instability and the demixing of particles and liquid. In
our scaling ($\varepsilon_{ll}=1$), demixing will occur for large
$\varepsilon_{nn}$ (at fixed $\varepsilon_{nl}$) or small
$\varepsilon_{nl}$ (at fixed $\varepsilon_{nn}$), i.e., for large
ratios $\varepsilon_{nn}$ / $\varepsilon_{nl}$.

Fig.~\ref{fig:fingers-enn} presents results for the average number of
fingers when changing $\varepsilon_{nn}$ in the range between $1.4$ and
$2.5$. Inspecting the figure one can distinguish four regions:
\begin{itemize}
\item[(i)] At small $\varepsilon_{nn}<1.95$ the particular strength of
  the interaction has nearly no influence; only a slight decrease of
  the finger number with increasing $\varepsilon_{nn}$ is discernible.
  We call this the 'transport regime' as only the transport
  properties play a role.
\item[(ii)] In an intermediate region $1.95<\varepsilon_{nn} <2.15$
  the finger number increases steeply by about a factor of
  three. Fingers emerge at the receding front and there is no demixing
  at all going on in the liquid bulk behind the front.  What we see is
  still a pure front instability that is, however, strongly influenced
  by the ratio of interaction parameters. We interpret this as a
  front-induced demixing that leads to fingering. In other words, the
  front itself acts as a 'nucleation site' for demixing. We call this a
  'demixing-induced front instability'. Note, however, that fingering
  still leads to a strongly anisotropic fingering pattern. Although
  the fingers break at some places they do in general not break up into
  small nanoparticle islands.
\item[(iii)] Increasing $\varepsilon_{nn}$ further above $2.15$, the
  finger number decreases again till about $<f>=15$ at about
  $\varepsilon_{nn}=2.25$.  This is mainly a geometric effect
  resulting from our one-dimensional finger counting routine. It
  reflects the fact that the fingers break up increasingly and the
  'dried in' nanoparticle pattern starts to look more and more
  isotropic. Demixing of particles and liquid occurs already in the
  the bulk liquid behind the front.
\item[(iv)] Beyond $\varepsilon_{nn}=2.15$ fluid and particles demix
  already in the bulk in a homogeneous manner, the remaining dried-in
  structure of nanoparticle islands is isotropic, and the emerging
  demixing length scale decreases with increasing $\varepsilon_{nn}$. The 'finger
  number' is one way to count the islands. Note, however, that it is
  not an adequate measure to study those isotropic structures (that
  are not of central interest here). The front resembles a liquid
  front receding inside a porous medium (formed by the nanoparticle
  islands).
\end{itemize}

Our interpretation of the ongoing physical processes is confirmed by
Fig.~\ref{fig:fingers-enl} showing the dependence of the average
number of fingers on the liquid-particle interaction strength
$\varepsilon_{nl}$ in the range between $1.0$ and $2.0$ for fixed
$\varepsilon_{nn}=2.0$. A similar sequence of regions (i) to (iv) is
found; this time however, with decreasing interaction parameter. This
agrees well with the considerations at the start of this section.

%%%%%%%%%%%%%%%%%%%%%%%%%%%%%%%%%%%%%%%%%%%%%%%%%%%%%%%%%%%%%%%%%%%%%%%%%%%%%%%
\section{Conclusions}
\mylab{sec:conc}
%%%%%%%%%%%%%%%%%%%%%%%%%%%%%%%%%%%%%%%%%%%%%%%%%%%%%%%%%%%%%%%%%%%%%%%%%%%%%%%
%
The present work has focused on pattern formation observed in various
experimental settings involving dewetting and drying nanofluids on
solid substrates. In addition to polygonal networks and spinodal-like
structures, branched structures have been reported in the experiments.
We have employed a kinetic Monte Carlo model to study pattern
formation driven by the interplay of evaporating solvent and diffusing
nanoparticles. A justification of the usage of a model that only
includes dewetting by evaporation but not by convective motion of the
solvent has been given based on experimental observations and scaling
considerations derived from a mesoscopic continuum model.

The model has first been used to analyze the influence of the
nanoparticles on the basic dewetting behavior, i.e., on spinodal
dewetting and also on dewetting by nucleation and growth of holes. It
has been found that the 'classical' hypothesis that the solute mainly
'decorates' and 'conserves' the volatile dewetting structures of the
solvent has to be amended in some important regions of parameter space. 
While it is true that the nanoparticles help to image the basic dewetting patterns (such as the
labyrinthine structure resulting from spinodal evaporative dewetting
or the random polygonal network resulting from nucleation and growth), they \emph{shift }the bimodal and
spinodal line in the chemical potential by a small amount that can be
estimated using a mean field argument. In consequence, they influence
the nucleation rate. Although in the present work we have not focused
 on this aspect, it should be further scrutinized in the future. 
More importantly, the nanoparticles strongly destabilize straight or
circular dewetting fronts in the nucleation regime. The
main body of the paper has entirely focused on a numerical study of
the underlying transverse front instability.

We have analyzed the dependence of the characteristics of the
branching patterns on the driving chemical potential, the mobility or
diffusivity of the nanoparticles, and their concentration. As a result
we have found that the mean number of fingers is almost independent of
the particle concentration. At a higher concentration one finds a very
similar number of slightly thicker fingers. The most influential
parameters are the particle diffusivity and the chemical potential. A
decreasing chemical potential [mobility] leads to denser [less dense]
finger patterns that develop earlier [later]. Note that a decreasing
chemical potential corresponds to an increasing driving force, i.e.,
to a larger mean velocity of the dewetting front.  As a result, we
have drawn the conclusion that \emph{the crucial factor determining
  the instability characteristics is the ratio of the timescales of
  the different transport processes}. In particular, a small ratio of
the velocity of the dewetting front and the mean diffusion velocity of
the particles renders the front more unstable. If the particle
diffusivity is too low the particles are 'collected' at the front. In
consequence, the front is slowed down, and an unstable stratification
of concentration evolves that triggers a self-optimization of the
front motion: The front expels particles into fingers that stay
behind. In this way it can maintain its velocity at a nearly constant
value. A similar effect was described for dewetting polymer films
\cite{ReSh01}. There, the dewetting front expels liquid from the
growing rim which collects the dewetted polymer and the liquid is left
behind in the form of deposited droplets. The present system, however,
allows one to also study the branching of the finger structure.

Beside the dependencies on particle density, mobility and driving
force we have also investigated the influence of the energetics of the
system, i.e., the dependence of the fingering on the interaction
strengths. We were especially interested in a possible coupling
between front motion and instability on the one hand and a
liquid-particle demixing on the other hand. On physical grounds we
have distinguished three different regimes: at small particle-particle
interaction strength or large particle-liquid interaction strength the
fingering is nearly independent of the interaction constants. That is, it is a
purely dynamic instability. We have called this the 'transport
regime'.  At large particle-particle interaction strength or small
particle-liquid interaction strength demixing of liquid and particles
occurs already in the bulk liquid. The retraction of the liquid front
then resembles the dewetting of a two-dimensional porous medium. At
intermediate interaction strengths, the finger density strongly
depends on the energetics of the system. We have found that particles
and liquid demix at the moving front rendering it transversally
unstable. The resulting length scale is determined by the dynamics and
the energetics of the system. We have called this a 'demixing-induced
front instability'.

In the light of the numerical findings presented here the experimental
results described above may be understood in terms of changes in
nanoparticle mobility resulting from differences in the length of the
carbon chains of the thiol ligands and the interpenetration of the
ligands of different particles. Particles with short thiol chains
diffuse faster and interpenetrate less. Therefore nearly no fingering
is observed for C$_5$ and C$_8$. Longer chains lead to slower
diffusion implying better developed fingers for C$_{10}$ and
C$_{12}$. It is likely that the decrease in fingering for
C$_{14}$-passivated nano\-particles results on the one hand from end
gauche defects in C$_{14}$ chains that tend to produce less pronounced
interdigitation than for C$_{12}$ thiols \cite{Paul08}. On the other
hand longer chain length will also lead to greater core-core
separation. This might drive $\epsilon_{nn}$ down and shift the system
from a front demixing-induced front instability towards the transport
regime where the finger number is lower.  The present experiments,
however, do not allow a direct comparison of finger numbers with the
simulations as the radial geometry of the 'normal' nucleated holes
does not facilitate a quantitative analysis. Future experiments that
study fingering using imposed straight dewetting fronts would be of
high interest.

We had based the quantitative analysis of the fingering on the
assumption of stationarity of the fingering process, i.e., on the
assumption that the main properties of the pattern do not depend on
the streamwise position. A detailed test of the assumption has shown
that it is valid as the fingering pattern is stationary after an
initial exponential growth in finger number that peaks at a maximal
value before settling on a slightly lower level via finger
coarsening. Following the transient phase the finger density fluctuates
around a stationary state. This observation not only validates the
quantification used but also confirms the hypothesis of
auto-optimization of the front velocity: the front collects particles
at a constant rate and leaves them behind in deposited fingers at the
same rate. However, not only is the overall rate stationary but so too are 
the spatial distribution of fingers and its dynamics including the
creation of new fingers (finger tips) and the
annihilation of fingers by merging (branching points).

We have checked that the results obtained are generic in the sense
that small changes in the model set-up do not change the behavior
qualitatively. In particular, we used an independently written code
for $1\times1$ nanoparticles to ensure that the nanoparticle size is
not a crucial element. The present work assumed $3\times3$
nanoparticles to allow for simple comparison with already published
results \cite{RRGB03,MBM04,Mart07}. To reach the same density of
fingers for $1\times1$ nanoparticles the mobility has to be smaller
than in the $3\times3$ case.

A crucial rule used in the algorithm is the contrast in the mobility
of the nanoparticles, i.e., the assumption that they do not
diffuse onto the dry substrate. If one lifts this rule and tries to
bound the nanoparticles stronger into the liquid by increasing the
particle-liquid interaction strength, i.e., one strongly increases the
energetic bias of the particles towards the retracting liquid, no
front instability has been found. This explains why Ising-type models
for grain growth that include mobile diffusing impurities, to our
knowledge have never reported instabilities of domain boundaries
caused by the diffusing impurities \cite{SrHa87,ViPl92,GJM95}. In all
these models the impurities diffuse equally well in the two phases.
They might, however, be energetically biased towards one of the phases
or towards the phase boundary. Our present results on evaporating
nanoparticle solutions suggest that unstable domain boundaries might
also be found in the ordering dynamics of a ferromagnetic system with
diffusing impurities under the influence of an external field if the
mobility of the impurities strongly depends on the phase.

Finally, we discus the limitations of the presently used kinetic Monte
Carlo model that is based on the assumptions that the dominant
processes can be captured in a two-dimensional setting neglecting the
film thickness of the evaporating film and that the only relevant
dynamics corresponds to particle diffusion and solvent evaporation.
It was argued above, based on the general structure of a thin film
model, that for film thicknesses $<10$~nm, the convective motion of
the solution can be neglected as compared to that due to
evaporation. We have also laid out in which way recent experiments
support this approximation. However, the experiments clearly show
effects that cannot be described employing the present model. These
effects are associated with the macroscopic dewetting front which
recedes before the ultrathin 'postcursor' film evaporatively dewets
(and thus produces the different types of structure described here and
elsewhere using variants of the two-dimensional kinetic Monte Carlo
model \cite{RRGB03,MBM04,Mart07,Paul08}). The macroscopic front can
also be unstable. That instability, however, obviously involves the
convective motion of the solution. On the other hand one notices in
the last three panels of Fig.~\ref{fig:nets} a bimodal network
structure. The small scale network one is able to describe with the
present model. The evolution of the large scale (mesoscopic) network,
however, occurs at larger film thicknesses and involves convective
motion of the solvent.  Different models are necessary to describe
those processes. Several mesoscale dynamic models such as , e.g., a
dynamic density functional theory for nanoparticle and liquid
densities and thin film models are possible candidates for such models
and will be investigated in the future.

%%%%%%%%%%%%%%%%%%%%%%%%%%%%%%%%%%%%%%%%%%%%%%%%%%%%%%%%%%%%%%%%%%%%%%%%
\section{Acknowledgments}
%%%%%%%%%%%%%%%%%%%%%%%%%%%%%%%%%%%%%%%%%%%%%%%%%%%%%%%%%%%%%%%%%%%%%%%%
%
We acknowledge support by the European Union via the FP6 Marie Curie
scheme [Grant MRTN-CT-2004005728 (PATTERNS)].

\end{document}